\documentclass[a4paper,11pt,sort&compress,number,5p]{elsarticle}
\usepackage[utf8]{inputenc}
\usepackage{lineno,hyperref}
\usepackage{pifont}
\usepackage{amsmath}
\usepackage{textcomp}
\usepackage{natbib}
\usepackage{fleqn}
\usepackage{graphicx}
\usepackage{txfonts}
\usepackage{multicol}
\usepackage{subfig}
\usepackage{multirow}
\usepackage{caption}
\usepackage{tabularx}
\usepackage{flushend}
\usepackage{adjustbox}
\usepackage{xcolor}
\usepackage{booktabs}
\usepackage{dcolumn}
\modulolinenumbers[5]
\journal{Journal of Alloys and Compounds}

\begin{document}
\emergencystretch 3em
\begin{frontmatter}

\title{Prediction of half-metallicity and spin-gapless semiconducting behavior in the new series of FeCr-based quaternary Heusler alloys: an ab initio study}
\author[1]{R.~Dhakal}
\ead{ramesh.dhakal91@gmail.com}

\author[1]{S.~Nepal\corref{cor1}}
\ead{sashinepal36@gmail.com}

\author[3]{I. Galanakis}
\ead{galanakis@upatras.gr}

\author[2]{R.~P.~Adhikari}
\ead{rajendra.adhikari@ku.edu.np}

\author[1]{G. C. Kaphle}
\ead{gopi.kaphle@cdp.tu.edu.np}

\cortext[cor1]{Corresponding author}

\address[1]{Central Department of Physics, Tribhuvan University, Kathmandu, Nepal}
\address[2]{Department of Physics, Kathmandu University, Dhulikhel, Nepal}
\address[3]{Department of Materials Science, School of Natural Sciences, University of Patras, GR-26504 Patras, Greece}
\begin{abstract}
This paper presents a detailed investigation of FeCr-based quaternary Heusler alloys. By using ultrasoft pseudopotential, electronic and magnetic properties of the compounds are studied within the framework of Density Functional Theory (DFT) by using the Quantum Espresso package. The thermodynamic, mechanical, and dynamical stability of the compounds is established through the comprehensive study of different mechanical parameters and phonon dispersion curves. The meticulous study of elastic parameters such as bulk, Young's, shear moduli, etc. is done to understand different mechanical properties. The FeCr-based compounds containing also Yttrium are studied to redress the contradictory electronic and magnetic properties observed in the literature. The interesting properties like half-metallicity and spin-gapless semiconducting (SGS) behavior are realized in the compounds under study.       
\end{abstract}


\end{frontmatter}

\section{Introduction}
In the last few decades, the rapid development and unprecedented progress in the field of spintronics has brought Heusler alloys to the center of attention garnering a lot of interest in the magnetism research community.   The desired high data processing speed, nonvolatility and low ohmic energy dissipation can be achieved by manipulating the spin and charge of the spintronics devices. The nearly complete spin polarization at the Fermi-level, high magnetic-ordering temperature and huge magnetoresistance of Heusler alloys make them potential candidates for spintronics devices like spin filters and spin valves. 
High \hyphenation{magnetic ordering} temperature makes these devices suitable for room temperature applications.\cite{spintronics,Hirohata,fong2013half}. The possibility of the implementation of these compounds in real spintronics devices and the major challenges  have been discussed in the literature \cite{HeuslerPropandGrowth,SpintronicsClaudia,HMAlloys-Galanakis-lec}. Since several members of Heusler compounds are known to possess properties like half-metallicity and spin gapless semiconducting behavior, the extensive research of these compounds is necessary to explore such novel properties in the new members of Heusler family. The half-metal combines the properties of both metal and semiconductor i.e. one spin channel shows usual metallic character while the other spin channel has a gap at the Fermi level demonstrating semiconducting behavior. In SGS, instead of a gap, the valence band maximum and conduction band minimum meet at the Fermi level and hence the electrons can be excited from the valence band to conduction band without or little energy.    

The strong half-metallic ferromagnets have been extensively studied in full and inverse structure with particular emphasis on Co and Mn-based compounds as they show coherent growth on semiconductors with very little disorder and have high curie temperature \cite{Galanakis2002,Galanakisinverse,Kandpalmainpaper}. In addition to the usual full and inverse Heusler structure, another class of Heusler compound with a 1:1:1:1 stoichiometry crystallizing in the so-called Y-structure having the LiMgPdSb as prototype material have drawn significant attention in the last decade, and yet in comparison to the other members of the family, these alloys are scarcely investigated.

In 2009, Dai \textit{et al.} reported the large half-metallic gap in CoFeMnSi\cite{dai2009new} and demonstrated that among three nonequivalent superstructures, the compound prefers the structure where half-metallicity occurs. In the following years, Ni and Co-based half-metallic ferromagnets were experimentally synthesized\cite{alijani2011quaternary,alijani2011electronic}; all the half-metals, except NiFeMnGa, were reported to have Curie temperature more than 550K rendering them potential candidates for room temperature applications. In this experimental study of Co-based quaternary Heusler alloys, CoFeMnSi was reported as half-metal corroborating the study of Dai \textit{et al}. However, while searching SGS on the series of quaternary Heusler compounds, Xu and their team predicted CoFeMnSi as Spin Gapless Semiconductor\cite{xu2013new}. Later, Bainsla \textit{et al.} carefully studied the transport and electrical properties of CoFeMnSi and found a larger deviation in its conductivity behavior from the usual metal and semiconductor\cite{bainsla2015spin}. The evidence of \hyphenation{temperature independent} carrier concentration and non-metallic electrical conductivity strongly suggested the SGS nature of the compound.

In 2013, using first-principles calculations,  Gao and collaborators studied CoFeCrZ compounds and predicted half-metallicity in these compounds \cite{gao2013large}. For CoFeCrSi and CoFeCrAl, the gap was significant and half-metallicity was robust against the lattice variation but CoFeCrAl lost its half-metallicity when on-site Coulomb electrostatic repulsion was taken into account. It is important to note that CoFeCrAl was reported as spin gapless semiconductor\cite{xu2013new}, nearly spin gapless\cite{ozdougan2013slater}, and was again contradicted by the experimental study\cite{BAINSLA201582}, supporting the half-metallic nature of the compound. However, CoFeCrGa has been experimentally verified as a spin gapless semiconductor up to the temperature of 250K \cite{bainsla2015origin}. From the above discussion, it is clear that SGS and half-metallicity are very difficult to distinguish in many cases and hence meticulous study of transport behavior and spin polarization measurements is necessary to find out the particular nature of the compounds. Not long ago, based on high throughput DFT calculation, Gao \textit{et al.} investigated a large number of quaternary Heusler compounds and found 70 stable spin gapless semiconductors \cite{gao2019high}. Recently, Aull and the team have studied the group of quaternary Heusler alloys and identified suitable SGSs and half-metals for reconfigurable magnetic tunnel diodes and transistors\cite{aull2019ab}.

In the present study, motivated by the exploration of different novel properties like SGS and half-metallicity in the quaternary Heusler compounds, we investigate the new series of FeCr-based quaternary Heusler alloys by using first-principle calculations. Since most of the work available in the literature focus on Co-based Heusler alloys, we attempt to study quaternary Heusler alloys considering different chemical compositions. To the best of our knowledge, four of the compounds studied in this paper have not been studied yet whereas relatively scant information is available about the other compounds. Lately, magnetic semiconducting behavior and half-metallicity have been predicted in FeRu-based quaternary Heusler alloys\cite{guo2018magnetic}. The Heusler alloys with rare-transition metal atoms such as Ru\cite{siteRu}, Rh\cite{ray2021strain}, Y\cite{rasul2019study,idrissi2020half} have found a place in recent literature.  Very recently, the study of few Fe-based compounds has appeared in literature\cite{SHAKIL2021157370} but preliminary convex hull distance screening of these compounds in Open Quantum Material Database (OQMD)\cite{saal2013materials}\cite{OQMDkirklin2015} shows that the value is greater than 0.20 eV/atom for the three compounds among four and hence further investigation is needed to establish the  thermodynamic stability. Their calculation reveals that one of the compounds, FeTiVIn, is mechanically unstable; the value of the convex hull distance for this compound is found to be 0.278 eV/atom. 

In this communication, we attempt to study the new series of FeCr-based quaternary Heusler alloys by considering the convex hull distance. We also study the FeCr-based  quaternary Heusler compounds containing Yttrium (Fe-Y) to sort out the observed differences in the literature. Since the quaternary Heusler compounds have become the new research area aiming to design compounds attractive for applications, various attempts have been made to study and synthesize them using neighboring d-elements from the periodic table . For example, CrVTiAl has been predicted to exist using ab-initio calculations \cite{galanakis2014crvtial} and then it was grown experimentally \cite{crvtial_exp}. Moreover, Fe, as well as Co, are usual constituents of Heusler compounds. This led to the study of the compounds where Fe is combined with Cr and V or Ti or Y which have fewer valence electrons than Cr. Furthermore, for most of these compounds, the results in the OQMD showed that they may be feasible experimentally adding more worth to our study. Hence, we investigate FeCrQZ (Q= Ti, V, Y and Z= Al, Ga, Ge, Si, In, Sn, As, Sb) compounds and FeMnCrSi by using a plane-wave pseudopotential based method. We start with the computational details and provide detailed discussion on different possible configurations of quaternary Heulser alloys. After discussing the mechanical properties, we focus on the electronic and magnetic properties of the FeCr-based compounds along with the explicit explanation on the obtained results of Fe-Y compounds. Finally, we conclude our study and present a brief summary.                                                                       
\section{Computational details}
Within the framework of density functional calculations,  we have implemented plane-wave pseudopotential based method as used in the Quantum Espresso package\cite{QE-2009,QE-2017}, to calculate the structural, electronic, and magnetic properties of the compounds under study.  For all the elements, we adopt the ultra-soft pseudopotentials. In order to approximate the exchange and correlation effect we exploit the Perdew-Burke-Ernzerhof generalized gradient approximation (PBE-GGA) functional\cite{perdew1996generalized}, with an energy cutoff of 100 Ry and a Monkhorst-Pack grid of $8\times8\times8$. For the non self-consistent field (nscf) calculation, a denser k-point sampling of $14\times14\times 14$ is considered and nice convergence has been accomplished. The Brillouin-Zone integration is performed by using the linear tetrahedron method\cite{lineartetrahedra} and total energy minimization technique is executed for structural optimization. For relaxation of structure, threshold for the convergence of force is set to 10$ ^{-4}  $ Ry/a.u. The threshold for the convergence of the total energy is set to $10^{-5}$ Ry and the criteria for the convergence of self-consistent cycle is  fixed at $10^{-8}$. Phonon dispersion curve is calculated using Phonopy code\cite{phonopy} in conjunction with Quatum Espresso code. For this, a 2x2x2 supercell is used.               
\begin{table}[h]
	\caption{Different possible site occupation of quaternary Heusler alloys. Here, Z denotes main group elements such as Al, Ga, Si, Sb , Sn, e.t.c and X, X$^\prime$, and Q are transition metals such as Fe, Mn, Cr, V e.t.c.  }
	\label{tab:table-I}
	\resizebox{0.48\textwidth}{!}{%
		\begin{tabular}{lcccc}
			\toprule[0.13em]
			\multirow{2}{*}{} & A       & B             & C             & D             \\ \cline{2-5}
			& (0,0,0) & ($\frac{1}{4}$,$\frac{1}{4}$,$\frac{1}{4}$) & ($\frac{1}{2}$,$\frac{1}{2}$,$\frac{1}{2}$) & ($\frac{3}{4}$,$\frac{3}{4}$,$\frac{3}{4}$) \\ \bottomrule[0.13em]
			Type-I            & Z       & X            & X\textquotesingle            & Q             \\ 
			Type-II           & Z       & X\textquotesingle           & X            & Q            \\ 
			Type-III          & X      & Z             & X\textquotesingle            & Q             \\ \bottomrule[0.13em]
		\end{tabular}%
	}
\end{table}

\section{Result and discussions}
The atomic order and arrangements largely dictate the electronic, magnetic, and transport  properties of the Heusler compounds. The proper understanding and comprehensive investigation of the crystal structure is necessary to unravel the structure-property dependence of Heusler compounds. It is a well-known fact that the Heusler structure is comprised of four interpenetrating face-centered cubic sublattice\cite{HeuslerPropandGrowth,SpintronicsClaudia}. Depending on the presence or absence of void, the crystal structure can be half-Heusler  or full Heusler and is denoted by XQZ (X, Q= transition metals, Z= main group elements) or X$_2$QZ respectively. The XQZ crystallizes in  the space group $F\bar{4}3m$(216) but X$_2$QZ can crystallize on either of the space groups, $F\bar{4}3m$(216) or $Fm\bar{3}m$(225), contingent on the arrangement of X and Q atoms.
\begin{table*}[ht!]
	\caption{Calculated values of total energy($ E_{tot} $), optimized lattice parameter($ a_{opt} $), and total magnetic moment($ m_{tot} $) of quaternary Heusler alloys for different possible structure. The type of structures are explained in table \ref{tab:table-I}. N$ _V $ is the number of valence electrons. }
	\label{tab:Table-II}
	\resizebox{1\textwidth}{!}{%
		\begin{tabular}{lcccc|lcccc|lcccc}
			\toprule[0.13em]
			\multirow{2}{*}{} &
			\multicolumn{1}{c}{\multirow{2}{*}{N$_V$}} &
			\multicolumn{1}{c}{\multirow{2}{*}{\begin{tabular}[c]{@{}c@{}}E$_{tot}$\\ (Rydberg)\end{tabular}}} &
			\multirow{2}{*}{\begin{tabular}[c]{@{}c@{}}a$_{opt}$\\ (\AA)\end{tabular}} &
			m$_{tot}$ &
			\multirow{2}{*}{} &
			\multicolumn{1}{c}{\multirow{2}{*}{N$_V$}} &
			\multicolumn{1}{c}{\multirow{2}{*}{\begin{tabular}[c]{@{}c@{}}E$_{tot}$\\ (Rydberg)\end{tabular}}} &
			\multirow{2}{*}{\begin{tabular}[c]{@{}c@{}}a$_{opt}$\\ (\AA)\end{tabular}} &
			m$_{tot}$ &
			\multirow{2}{*}{} &
			\multicolumn{1}{c}{\multirow{2}{*}{N$_V$}} &
			\multicolumn{1}{c}{\multirow{2}{*}{\begin{tabular}[c]{@{}c@{}}E$_{tot}$\\ (Rydberg)\end{tabular}}} &
			\multirow{2}{*}{\begin{tabular}[c]{@{}c@{}}a$_{opt}$\\ (\AA)\end{tabular}} &
			m$_{tot}$ \\
			&
			\multicolumn{1}{c}{} &
			\multicolumn{1}{c}{} &
			&
			\textbf{($\mu_B$)} &
			&
			\multicolumn{1}{c}{} &
			\multicolumn{1}{c}{} &
			&
			($\mu_B$) &
			&
			\multicolumn{1}{c}{} &
			\multicolumn{1}{c}{} &
			&
			\multicolumn{1}{c}{($\mu_B$)} \\ \bottomrule[0.13em]
			\textbf{FeCrTiAl} &
			&
			&
			&
			&
			\textbf{FeCrVGe} &
			\textbf{} &
			\textbf{} &
			\textbf{} &
			\textbf{} &
			\textbf{FeCrYSi} &
			\multicolumn{1}{l}{\textbf{}} &
			\multicolumn{1}{l}{\textbf{}} &
			\multicolumn{1}{l}{\textbf{}} &
			\multicolumn{1}{l}{\textbf{}} \\ 
			Type-I &
			\multirow{3}{*}{21} &
			-547.60388 &
			5.9820 &
			-2.29 &
			Type-I &
			\multirow{3}{*}{23} &
			-580.15029 &
			5.8598 &
			-0.99 &
			Type-I &
			\multirow{3}{*}{21} &
			-516.46757 &
			6.4541 &
			2.68 \\ 
			Type-II &
			&
			-547.62206 &
			6.0550 &
			-3.35 &
			Type-II &
			&
			-580.16522 &
			5.9162 &
			-0.80 &
			Type-II &
			&
			-516.50092 &
			6.4840 &
			1.10 \\ 
			Type-III &
			&
			-547.68596 &
			5.9771 &
			-2.99 &
			Type-III &
			&
			-580.19664 &
			5.8070 &
			-0.99 &
			Type-III &
			&
			-516.53859 &
			6.2360 &
			3.00 \\ \hline
			\textbf{FeCrTiGa} &
			&
			&
			&
			&
			\textbf{FeCrVSi} &
			\textbf{} &
			\textbf{} &
			\textbf{} &
			\textbf{} &
			\textbf{FeCrYGe} &
			\multicolumn{1}{l}{\textbf{}} &
			\multicolumn{1}{l}{\textbf{}} &
			\multicolumn{1}{l}{\textbf{}} &
			\multicolumn{1}{l}{\textbf{}} \\ 
			Type-I &
			\multirow{3}{*}{21} &
			-713.59824 &
			5.9851 &
			-2.75 &
			Type-I &
			\multirow{3}{*}{23} &
			-578.64443 &
			5.7322 &
			-1.12 &
			Type-I &
			\multirow{3}{*}{21} &
			-518.04148 &
			6.5795 &
			2.70 \\ 
			Type-II &
			&
			-713.62479 &
			6.0398 &
			-3.35 &
			Type-II &
			&
			-578.64650 &
			5.8047 &
			-1.03 &
			Type-II &
			&
			-518.07467 &
			6.5816 &
			1.07 \\ 
			Type-III &
			&
			-713.66628 &
			5.9671 &
			-2.99 &
			Type-III &
			&
			-578.69780 &
			5.7081 &
			-0.99 &
			Type-III &
			&
			-518.08979 &
			6.3385 &
			3.00 \\ \hline
			\textbf{FeCrTiGe} &
			&
			&
			&
			&
			\textbf{FeMnCrSi} &
			\multicolumn{1}{l}{\textbf{}} &
			\multicolumn{1}{l}{\textbf{}} &
			\multicolumn{1}{l}{\textbf{}} &
			\multicolumn{1}{l|}{\textbf{}} &
			\textbf{FeCrYSn} &
			\multicolumn{1}{l}{\textbf{}} &
			\multicolumn{1}{l}{\textbf{}} &
			\multicolumn{1}{l}{\textbf{}} &
			\multicolumn{1}{l}{\textbf{}} \\ 
			Type-I &
			\multirow{3}{*}{22} &
			-554.34622 &
			6.0115 &
			-3.67 &
			Type-I &
			\multirow{3}{*}{25} &
			-646.95709 &
			5.6904 &
			1.01 &
			Type-I &
			\multirow{3}{*}{21} &
			-669.27423 &
			6.8383 &
			2.59 \\ 
			Type-II &
			&
			-554.38815 &
			6.0446 &
			-4.02 &
			Type-II &
			&
			-646.94332 &
			5.6832 &
			0.96 &
			Type-II &
			&
			-669.29768 &
			6.8479 &
			1.32 \\ 
			Type-III &
			&
			-554.41529 &
			5.9152 &
			-1.99 &
			Type-III &
			&
			-646.97016 &
			5.5960 &
			0.99 &
			Type-III &
			&
			-669.33858 &
			6.6077 &
			3.00 \\ \hline
			\textbf{FeCrTiSi} &
			&
			&
			&
			&
			\textbf{FeCrYAl} &
			\multicolumn{1}{l}{\textbf{}} &
			\multicolumn{1}{l}{\textbf{}} &
			\multicolumn{1}{l}{\textbf{}} &
			\multicolumn{1}{l|}{\textbf{}} &
			\textbf{FeCrYAs} &
			\multicolumn{1}{l}{\textbf{}} &
			\multicolumn{1}{l}{\textbf{}} &
			\multicolumn{1}{l}{\textbf{}} &
			\multicolumn{1}{l}{\textbf{}} \\ 
			Type-I &
			\multirow{3}{*}{22} &
			-552.81886 &
			5.8454 &
			-1.71 &
			Type-I &
			\multirow{3}{*}{20} &
			-511.28024 &
			6.5831 &
			1.94 &
			Type-I &
			\multirow{3}{*}{22} &
			-526.70470 &
			6.6305 &
			2.28 \\ 
			Type-II &
			&
			-552.85170 &
			5.9339 &
			-3.65 &
			Type-II &
			&
			-511.29871 &
			6.6000 &
			0.49 &
			Type-II &
			&
			-526.75589 &
			6.5369 &
			5.54 \\ 
			Type-III &
			&
			-552.90210 &
			5.8127 &
			-1.99 &
			Type-III &
			&
			-511.32948 &
			6.4649 &
			2.00 &
			Type-III &
			&
			-526.72829 &
			6.3316 &
			4.00 \\ \hline
			\textbf{FeCrVAl} &
			&
			&
			&
			&
			\textbf{FeCrYGa} &
			\multicolumn{1}{l}{\textbf{}} &
			\multicolumn{1}{l}{\textbf{}} &
			\multicolumn{1}{l}{\textbf{}} &
			\multicolumn{1}{l|}{\textbf{}} &
			\textbf{FeCrYSb} &
			\multicolumn{1}{l}{\textbf{}} &
			\multicolumn{1}{l}{\textbf{}} &
			\multicolumn{1}{l}{\textbf{}} &
			\multicolumn{1}{l}{\textbf{}} \\ 
			Type-I &
			\multirow{3}{*}{22} &
			-573.42612 &
			5.8718 &
			-2.00 &
			Type-I &
			\multirow{3}{*}{20} &
			-677.29614 &
			6.5805 &
			1.94 &
			Type-I &
			\multirow{3}{*}{22} &
			-526.12519 &
			6.8635 &
			7.02 \\ 
			Type-II &
			&
			-573.40401 &
			5.9739 &
			-3.91 &
			Type-II &
			&
			-677.31859 &
			6.5793 &
			0.58 &
			Type-II &
			&
			-526.12333 &
			6.7736 &
			6.59 \\ 
			Type-III &
			&
			-573.46867 &
			5.8481 &
			-2.00 &
			Type-III &
			&
			-677.32187 &
			6.4490 &
			2.00 &
			Type-III &
			&
			-526.18111 &
			6.5570 &
			4.00 \\ \hline
			\textbf{FeCrVGa} &
			\textbf{} &
			\textbf{} &
			\textbf{} &
			\textbf{} &
			\textbf{FeCrYIn} &
			\multicolumn{1}{l}{\textbf{}} &
			\multicolumn{1}{l}{\textbf{}} &
			\multicolumn{1}{l}{\textbf{}} &
			\multicolumn{1}{l|}{\textbf{}} &
			&
			\multicolumn{1}{l}{} &
			\multicolumn{1}{l}{} &
			\multicolumn{1}{l}{} &
			\multicolumn{1}{l}{} \\ 
			Type-I &
			\multirow{3}{*}{22} &
			-739.40261 &
			5.8793 &
			-1.93 &
			Type-I &
			\multirow{3}{*}{20} &
			-646.47973 &
			6.8426 &
			1.85 &
			&
			\multicolumn{1}{l}{} &
			\multicolumn{1}{l}{} &
			\multicolumn{1}{l}{} &
			\multicolumn{1}{l}{} \\ 
			Type-II &
			&
			-739.39998 &
			5.9724 &
			-3.91 &
			Type-II &
			&
			-646.49407 &
			6.8367 &
			0.86 &
			\multicolumn{1}{c}{} &
			\multicolumn{1}{l}{} &
			\multicolumn{1}{l}{} &
			\multicolumn{1}{l}{} &
			\multicolumn{1}{l}{} \\ 
			Type-III &
			&
			-739.44140 &
			5.8513 &
			-1.99 &
			Type-III &
			&
			-646.51983 &
			6.7165 &
			2.01 &
			\multicolumn{1}{c}{} &
			\multicolumn{1}{l}{} &
			\multicolumn{1}{l}{} &
			\multicolumn{1}{l}{} &
			\multicolumn{1}{l}{} \\ \bottomrule[0.13em]
		\end{tabular}%
	}
\end{table*}
The respective position of X and Q atoms in the lattice comes from the electronegativity arguments; when the valence of X atoms is larger than the valence of Q atoms, the compound crystallizes in a full Heusler structure(225), where the arrangement of the atoms along the diagonal is X-Q-X-Z. When the valence of X atoms is smaller than Q atoms, the sequence is altered to X-X-Q-Z and hence the compound crystallizes in the inverse Heusler structure with space group 216. The structure becomes quaternary(XX\textquotesingle\,QZ) when among two similar X atoms one is replaced by a new transition metal atom X\textquotesingle\,  and the crystal with symmetry F$\bar{4}$3m(216)  is reproduced. Theoretically, the four atoms of quaternary Heusler alloys can occupy different  Wyckoff positions generating three nonequivalent superstructures \cite{alijani2011electronic}. As shown in the table\ref{tab:table-I}, by switching the Wyckoff positions of three of the atoms one can get type-I, type-II, and type-III structures for each of the compounds. The coordinates for A, B, C and D sublattices are (0, 0, 0),  ($\frac{1}{4}$, $\frac{1}{4}$,  $\frac{1}{4}$), ($\frac{1}{2}$, $\frac{1}{2}$, $\frac{1}{2}$) and ($\frac{3}{4}$, $\frac{3}{4}$, $\frac{3}{4}$) respectively in Wyckoff coordinates.

The search for novel materials with desirable properties starts by identifying the promising candidates which can be possibly synthesized. Among many constraints that one should look upon for the probabilistic construction of the materials, an important one is the formation energy of the compounds. The formation energy, which is the difference in energy between the total energy of the compound in bulk forms and the sum of the energy of individual atoms in the elemental phase, is the necessary condition to synthesize the materials as its negative value suggests that the compound is stable against its characteristic elements at the ground state. However, having negative formation energy is not enough to assure the stability of the particular structure over other similar phases at that composition. Thus, one should look after the convex-hull, to find out whether the given phase is thermodynamically stable and is likely to be  fabricated. Convex-hull can be defined as the energy difference between the studied phase and the most stable phase combination keeping the stoichiometry. For a detailed understanding of stability criterion and convex-hull, in the case of Heusler alloys, readers are referred to \cite{gao2019high, ma2018computational}. 

The \href{http://oqmd.org/}{OQMD} is a searchable database of DFT calculated thermodynamic and structural properties of 815654 materials, created in \href{http://wolverton.northwestern.edu}{Chris Wolverton}’s group in Northwestern University. They have employed the Vienna Ab-initio Simulation Package (\href{https://www.vasp.at/}{VASP})\cite{vasp}, version 5.3.2 for the calculations in conjunction with the generalized gradient approximation (GGA) to the exchange and correlation potential as parametrized by Perdew, Burke, and Ernzerhof (PBE)\cite{perdew1996generalized}. Concerning the choice of pseudopotentials they have employed the projected augmented wave (PAW)\cite{PAW} method. They have started from several initial configurations in order, not only to accurately describe the phase diagram for a certain combination of chemical elements, but also to study and include in their database all possible metastable and not stable lattice structures. In our work by using OQMD, we first collect the list of possible FeCr-based quaternary Heusler alloys whose convex Hull distance is less than 200 meV per unit cell. We have set this threshold because according to  reference\cite{wu2013first} most metastable phases correspond to 36 meV/atom energy
\begin{table*}[ht!]
	\caption{Hull distance($ E_{hull} $), formation energy($ E_{form} $), optimized lattice parameter($ a_{opt} $), atom-resolved ($ m_{element}; element = X, X^\prime, Q, and\,Z $) and total spin magnetic moment($ m_{tot} $) of Fe-based quaternary Heusler alloys. Here, the value of hull distance and formation energy is taken from OQMD databse \cite{saal2013materials}\cite{OQMDkirklin2015}. N$ _V $ is the number of valence electrons.}
	\label{tab:Table-III}
	\resizebox{\textwidth}{!}{%
		\begin{tabular}{lccccccccc}
			\toprule[0.13em]
			\multirow{2}{*}{\begin{tabular}[l]{@{}l@{}}Compounds\\ (XX$^\prime$QZ)\end{tabular}} &
			\multicolumn{1}{l}{\multirow{2}{*}{N$_V$}} &
			\multirow{2}{*}{\begin{tabular}[c]{@{}c@{}}E$_{hull}$\\ (eV/atom)\end{tabular}} &
			\multirow{2}{*}{\begin{tabular}[c]{@{}c@{}}E$_{form}$\\ (eV/atom)\end{tabular}} &
			\multirow{2}{*}{\begin{tabular}[c]{@{}c@{}}a$_{opt}$\\ (\AA)\end{tabular}} &
			\multicolumn{5}{c}{Magnetic Moments ($\mu_B$)} \\ \cline{6-10} 
			& \multicolumn{1}{l}{} &       &        &        & m$_{tot}$ & m$_{X}$ & m$_{X^\prime}$ & m$_{Q}$ & m$_{Z}$\\ \bottomrule[0.13em]
			FeCrTiAl & 21                    & 0.036 & -0.310 & 5.9771 & -2.99              & -0.6833 & -2.4286 & 0.2576  & 0.0285  \\ 
			FeCrTiGa & 21                    & 0.058 & -0.261 & 5.9671 & -2.99              & -0.6769 & -2.4584 & 0.2685  & 0.0307  \\ 
			FeCrTiGe & 22                    & 0.149 & -0.274 & 5.9152 & -1.99              & -0.6145 & -1.5970 & 0.2762  & 0.0315  \\ 
			FeCrTiSi & 22                    & 0.143 & -0.446 & 5.8127 & -1.99              & -0.6304 & -1.4550 & 0.1920  & 0.0265  \\ 
			FeCrVAl  & 22                    & 0.072 & -0.211 & 5.8481 & -2.00              & -1.0580 & -1.6310 & 0.7156  & 0.0077  \\ 
			FeCrVGa  & 22                    & 0.074 & -0.135 & 5.8513 & -1.99              & -1.0341 & -1.6946 & 0.7507  & 0.0127  \\ 
			FeCrVGe  & 23                    & 0.076 & -0.171 & 5.8070 & -0.99              & -0.6044 & -0.8389 & 0.4384  & 0.0119  \\ 
			FeCrVSi  & 23                    & 0.053 & -0.391 & 5.7081 & -0.99              & -0.6091 & -0.7417 & 0.3606  & 0.0088  \\ 
			FeMnCrSi & 25                    & 0.042 & -0.314 & 5.5960 & 0.99               & 0.1466  & -0.3082 & 1.0866  & -0.0058  \\\hline 
			FeCrYAl  & 20                    & 0.368 & 0.142  & 6.4649 & 2.00               & -1.6780 & 3.2899  & 0.0797  & -0.0168 \\ 
			FeCrYGa  & 20                    & 0.497 & 0.164  & 6.4490 & 2.00               & -1.7645 & 3.3459  & 0.1014  & -0.0254 \\ 
			FeCrYIn  & 20                    & 0.500 & 0.247  & 6.7165 & 2.01               & -2.1288 & 3.6135  & 0.1464  & -0.0150 \\ 
			FeCrYSi  & 21                    & 0.488 & 0.006  & 6.2360 & 3.00               & -0.0396 & 2.8399  & 0.0155  & -0.0533 \\ 
			FeCrYGe  & 21                    & 0.499 & 0.051  & 6.3385 & 3.00               & -0.2984 & 3.0559  & 0.0272  & -0.0495 \\ 
			FeCrYSn  & 21                    & 0.388 & 0.015  & 6.6077 & 3.00               & -0.7191 & 3.3788  & 0.0559  & -0.0310 \\ 
			FeCrYAs  & 22                    & 0.845 & 0.067  & 6.5369 & 5.54               & 2.4004  & 2.9766  & -0.0589 & -0.0271 \\ 
			FeCrYSb  & 22                    & 0.667 & 0.083  & 6.5570 & 4.00               & 1.1646  & 3.0419  & -0.1287 & -0.0587 \\ \bottomrule[0.13em]
		\end{tabular}%
	}
\end{table*}
distance from the ground state. Authors in that reference have used energy data from the inorganic crystal structure database (ICSD)\cite{icsd} comparing the energy difference between metastable and stable phases of the compounds in that database. In our case, we have four atoms per unit cell, and thus $4\times36=144$ meV. We can safely assume that within a convex Hull distance of 200 meV per unit cell, one can find almost all metastable phases provided in the OQMD database. Here, we report only those FeCr-based quaternary Heusler compounds whose properties we find interesting to investigate. To achieve the equilibrium lattice structures of the compounds under study, the lattice parameter is optimized by considering the different initial magnetic configuration for each type of atomic arrangement. Among the three structures, the optimized calculation shows the lowest energy for the type-III structure for all the compounds. The computed value of the total energy along with the corresponding lattice constant is shown in table \ref{tab:Table-II}. The type-III structure is the most stable structure for all compounds, except for FeCrYAs for which the type-II structure is stable. We perform all the calculations henceforward by considering the respective stable configuration. The dynamical stability of the compounds in table \ref{tab:tableElastic} is also confirmed by calculating the phonon dispersion curves. The presence of negative frequencies (imaginary part) in phonon dispersion curve indicates dynamical instability. We found no such negative frequencies in phonon dispersion curves (not shown here) of the compounds, thus confirming their dynamical stability.

\subsection{Mechanical properties}
In the previous section, we discussed the thermodynamic stability of the compounds under study. In this section, to further establish the stability of the compounds, we calculated the second-order elastic constants, C$ _{ij} $, using \textbf{ElaStic} code \cite{elasticTool}. All the compounds under study have a cubic structure. For cubic crystals, there are three independent second-order elastic constants--C$_{11}$, C$ _{12} $, and C$ _{44} $. The cubic crystals which satisfy the Born-Huang stability criteria \cite{born_huang_1956} ,given as
\begin{equation}
\label{eqn:born-huang_criterion}
C_{11} + 2C_{12} > 0, \,\,\, C_{44}>0,\,\,\,C_{11}>C_{12}
\end{equation}
are considered mechanically stable. The elastic constants of the compounds under study are reported in table \ref{tab:tableElastic} and it is found that in all case the Born-Huang stability criteria (\ref{eqn:born-huang_criterion}) is satisfied. Hence, the compounds in table \ref{tab:tableElastic} are mechanically stable.

\begin{table*}[h!]
		\caption{Calculated values of elastic constants ($ C_{ij} $ in GPa), Bulk modulus ($ B $ in GPa), Shear modulus ($ G $ in GPa), Young's modulus (E in GPa), Pugh's ratio ($ \Bbbk = B/G$), Poisson's ration ($ \nu $), Cauchy pressure ($ C_p $) and Anisotropic factor ($ A_e $). Data of $ C_{ij} $ values are taken from literature \cite{diamond,elements} for Diamond, Iridium, Silver, Platinum, and Gold and other elastic properties are calculated using those values. This data are included for comparison only.}
	\label{tab:tableElastic}
	\resizebox{\textwidth}{!}{%
		\begin{tabular}{lccccccccccc}
			\toprule[0.13em]
			Compounds & $C_{11}$ & $C_{12}$ & $C_{44}$ & $B$    & $G$    & E      & $\Bbbk=B/G$ & G/B                       & $\nu$ & $C_p$  & $A_e$ \\ \bottomrule[0.13em]
			FeCrTiAl  & 264.2    & 109.0    & 107.3    & 160.69 & 94.25  & 236.51 & 1.70        & 0.59                      & 0.25  & 1.70   & 1.38  \\ 
			FeCrTiGa  & 253.7    & 121.9    & 106.0    & 165.82 & 87.58  & 223.41 & 1.89        & 0.53                      & 0.28  & 15.90  & 1.61  \\ 
			FeCrTiGe  & 253.0    & 144.2    & 106.0    & 180.45 & 81.13  & 211.67 & 2.22        & 0.45                      & 0.30  & 38.20  & 1.95  \\ 
			FeCrTiSi  & 271.0    & 168.0    & 114.2    & 202.31 & 82.96  & 218.95 & 2.44        & 0.41                      & 0.32  & 53.80  & 2.22  \\ 
			FeCrVAl   & 277.8    & 133.2    & 117.2    & 181.42 & 96.55  & 246.01 & 1.88        & 0.53                      & 0.27  & 16.00  & 1.62  \\ 
			FeCrVGa   & 243.7    & 149.4    & 111.6    & 180.86 & 78.98  & 206.83 & 2.29        & 0.44                      & 0.31  & 37.80  & 2.37  \\ 
			FeCrVGe   & 315.2    & 167.4    & 120.6    & 216.71 & 99.11  & 258.00 & 2.19        & 0.46                      & 0.30  & 46.80  & 1.63  \\ 
			FeCrVSi   & 368.2    & 169.1    & 136.7    & 235.44 & 120.40 & 308.60 & 1.96        & 0.51                      & 0.28  & 32.40  & 1.37  \\ 
			FeMnCrSi  & 412.6    & 167.9    & 132.7    & 249.44 & 128.45 & 328.89 & 1.94        & 0.51                      & 0.28  & 35.20  & 1.08  \\ \hline
			Diamond$^a$ & 950.0 & 390.0 & 420.0 & 576.67 & 357.00 & 887.80 & 1.62 & \multicolumn{1}{c}{0.62} & 0.24 & -30.00 & 1.50 \\ 
			Irridium$^b$    & 580.0 & 260.0 & 270.0 & 366.67 & 218.88 & 547.67 & 1.68 & \multicolumn{1}{c}{0.60} & 0.25 & -10.00 & 1.69 \\
			Silver$^b$    & 122.0    & 92.0     & 45.5     & 102.00 & 29.20  & 79.96  & 3.49        & \multicolumn{1}{c}{0.29} & 0.37  & 46.50  & 3.03    \\
			Platinum$^b$    & 347.0 & 251.0 & 76.5  & 283.00 & 63.46  & 177.14 & 4.46 & \multicolumn{1}{c}{0.22} & 0.40 & 174.50 & 1.59 \\ 
			Gold$^b$      & 192.9    & 163.8    & 41.5     & 173.50 & 27.28  & 77.76  & 6.36        & \multicolumn{1}{c}{0.16} & 0.43  & 122.30 & 2.85    \\ \bottomrule[0.13em]
		\end{tabular}%
	}
References:\\
$ ^a $ref \cite{diamond}, $^b$ref \cite{elements}
\end{table*}

Using the calculated elastic constants, we can evaluate different elastic parameters for the crystals such as Young's, shear and bulk moduli, anisotropy factors, etc.
Polycrystalline materials are composed of many single crystal grains which are oriented randomly. Bulk (B) and Shear (G) moduli can be used to describe a isotropic system completely. These moduli can be evaluated by averaging over second-order elastic stiffness (C$_{ij}$) or elastic compliance (S$_{ij}$). We can calculate ab-initio polycrystalline elastic moduli by first calculating single-crystal elastic stiffness (C$ _{ij} $) and/or elastic compliance (S$ _{ij} $) and then transforming these data to macroscopic physical quantities by suitable averaging methods. The Voigt-Ruess-Hill averaging methods are the most popular and can be used to calculate polycrystalline elastic moduli. 
A uniform strain is assumed in Voigt \cite{voigt1910lehrbuch} method and it uses elastic stiffness C$ _{ij} $ to compute elastic moluli while uniform stress is assumed in Ruess \cite{ruess1929} method which uses elastic compliance S$ _{ij} $ to compute elastic moduli. For cubic systems, the Bulk modulus by Voigt and Ruess approaches ($ B_V\,\&\,B_R $) are equal and are given by 
\begin{equation}
B=B_V=B_R=\frac{C_{11}+2C_{12}}{3}=\frac{1}{3(S_{11}+2S_{12})}
\end{equation}
For a cubic system, the Shear modulus according to Voigt and Ruess are given by 
\begin{equation}
G_V = \frac{C_{11}-C_{12} + 3C_{44}}{5}
\end{equation}
and
\begin{equation}
\begin{aligned}
G_R=\frac{5}{4(S_{11}-S_{12})+3S_{44}}\\
=\frac{5C_{44}(C_{11} - C_{12})}{3(C_{11}-C_{12})+C_{44}}
\end{aligned}
\end{equation}
In Hill's averaging approach \cite{Hill_1952,HILL1963}, the Voigt's and Reuss's elastic moduli are taken as upper and lower bound and the bulk and shear moduli are determined as 
\begin{equation}
B_H= \frac{B_V+B_R}{2}=B
\end{equation}
and
\begin{equation}
G_H = G = \frac{G_V+G_R}{2}
\end{equation}
Pugh ratio ($ \Bbbk$), Young's modulus (E), and Poisson's ratio ($\nu$) are determined using the averaged bulk and shear moduli, B \& G as
\begin{equation}
\Bbbk = \frac{B}{G}
\end{equation} 
\begin{equation}
E = 2G(1+\nu)
\end{equation}
\begin{equation}\label{eqn:poissonandpugh}
\nu = \frac{3B-2G}{2(3B+G)} = \frac{3\Bbbk - 2 }{6\Bbbk +2}
\end{equation}
\begin{figure}[h!]
	\centering
	\includegraphics[width=1\linewidth]{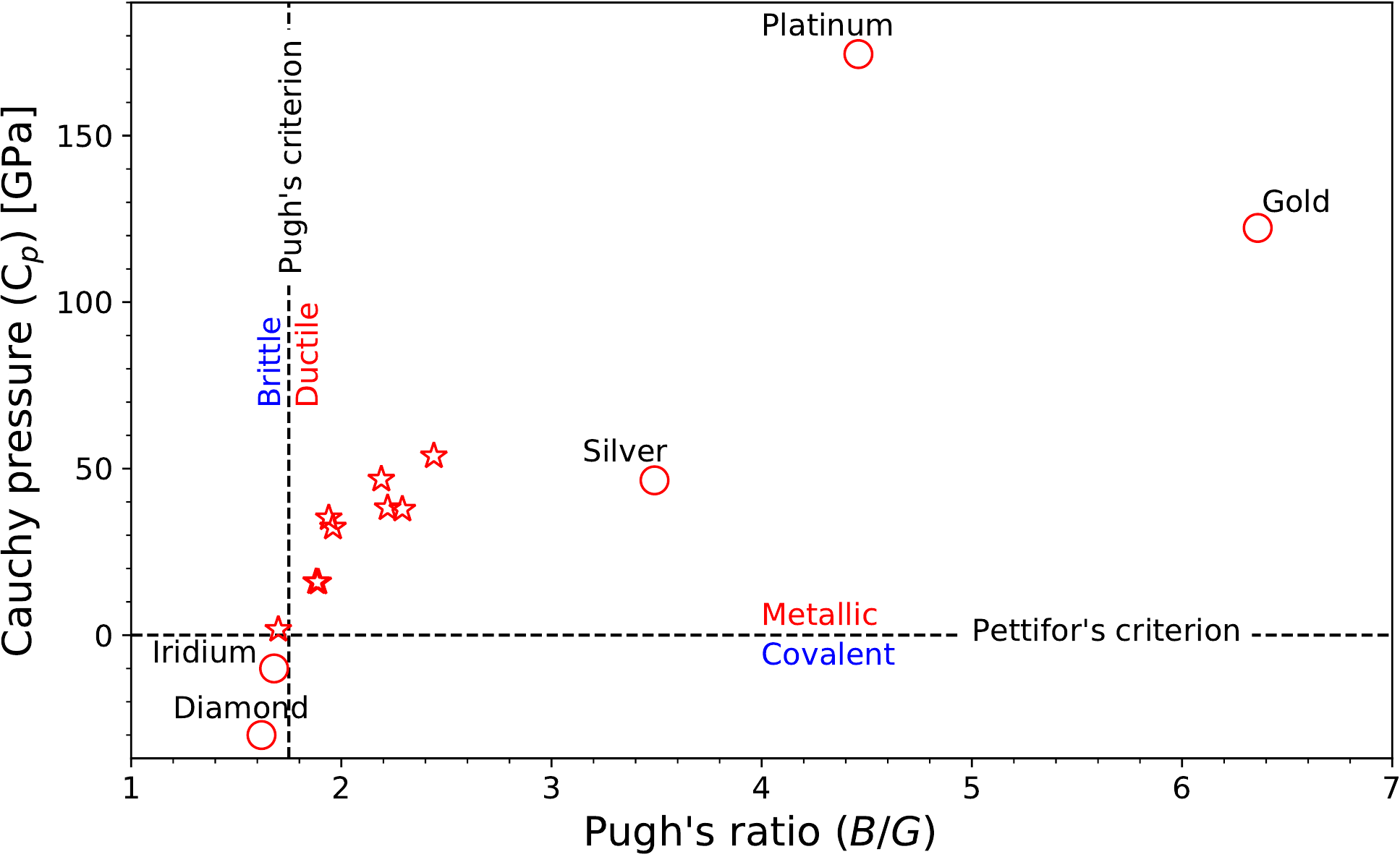}
	\caption{Plot of Cauchy pressure ($ C_p $) vs Pugh's ratio ($ B/G $). Horizontal and vertical dashed line corresponds to the Pettifor's and Pugh's criterion as explained above in the text. Data of Diamond (which is known to be the hardest) and Gold (which is known to be the most ductile), and also of Iridium, Silver, Platinum is presented for better comparison.}
	\label{fig:ductility}
\end{figure}

We can characterize material as ductile or brittle. In some applications, brittleness of material is desired for example, if machinability of material is required. While in various applications malleability (ductility) is desired, for instance, if the material is needed to be drawn in the form of wire. Ductility of materials infers the presence of metallic bonding while brittleness infers the ionic or covalent character of bonding. Pugh ratio ($ \Bbbk $) and Poisson's ratio ($\nu$) can be used as a parameter to characterize the material as ductile or brittle. The calculated Pugh ratio and Poisson's ratio for the compounds under study are reported in table \ref{tab:tableElastic}. Pugh's criterion dictates the critical value for Pugh's ratio ($\Bbbk_c$), that separates brittle and ductile materials, is 1.75 \cite{elastic_Iyigor_BGValue,elastic_Seh,elastic_claudia}. If $ \Bbbk > \Bbbk_c $ for a given material then it is characterized as ductile and if $ \Bbbk < \Bbbk_c $ then the material is characterized as brittle. Obtained values of Pugh's ratio is greater than the critical value 1.75 except for FeCrTiAl, inferring that the compounds are ductile in nature except FeCrTiAl. This is further corroborated by calculating the Cauchy pressure, $ C_p=C_{12}-C_{44} $. According to Pettifor's \cite{pettifor} criterion, a positive Cauchy pressure indicates metallic bonding while its negative value indicates covalent bonding (directional bonding). Thus, a material with positive Cauchy pressure can be considered as ductile and that with negative Cauchy pressure as brittle. All compounds in table \ref{tab:tableElastic} have a positive value of Cauchy pressure and therefore inferring compounds to be ductile. Form figure \ref{fig:ductility} one can see that for FeCrTiAl, the behavior is not that simple. One must note that characterization of ductility and brittleness is not always straight forward. Comparing with data of diamond, iridium, silver, platinum, and gold, one can safely say that the behavior of FeCrTiAl lies in the borderline of ductility and brittleness.  

The Elastic anisotropy factor or Zener ratio ($A_e$) is another important parameter for description of mechanical stability and for the cubic crystal, it is given by
\begin{equation}
A_e=\frac{2C_{44}}{C_{11}-C_{12}}
\end{equation} 
The value of $A_e $ is 1 for isotropic material. Materials with high anisotropy tends to deviate from the cubic structure. Materials with $ A_e< 0 $ i.e negative anisotropy violates at least one Born-Huang criteria \ref{eqn:born-huang_criterion} and hence are mechanically unstable. All of the compounds in table \ref{tab:tableElastic} have a value of $ A_e $ greater than unity suggesting anisotropy. Among the given compounds, FeCrTiSi and FeCrVGa are more anisotropic than other compounds. 

\subsection{Slater Pauling rule and the hybridization scheme}
Heusler compounds are known to follow the Slater-Pauling rule. This rule relates the observed total spin
magnetic moment of the Heusler compounds with the valence electron concentration on the basis of the orbitals hybridization scheme. The total number of states in
the spin-down channel is predetermined because of which
the rise in the valence electron concentration results in the
increase of the number of electrons in the spin-up channel.
This phenomenon enables us to predict the total spin magnetic moment of the compounds since the magnetic moments increase proportionally with the increase in the total number of valence electrons.
\begin{figure}[h!]
	\centering
	\includegraphics[width=1\linewidth]{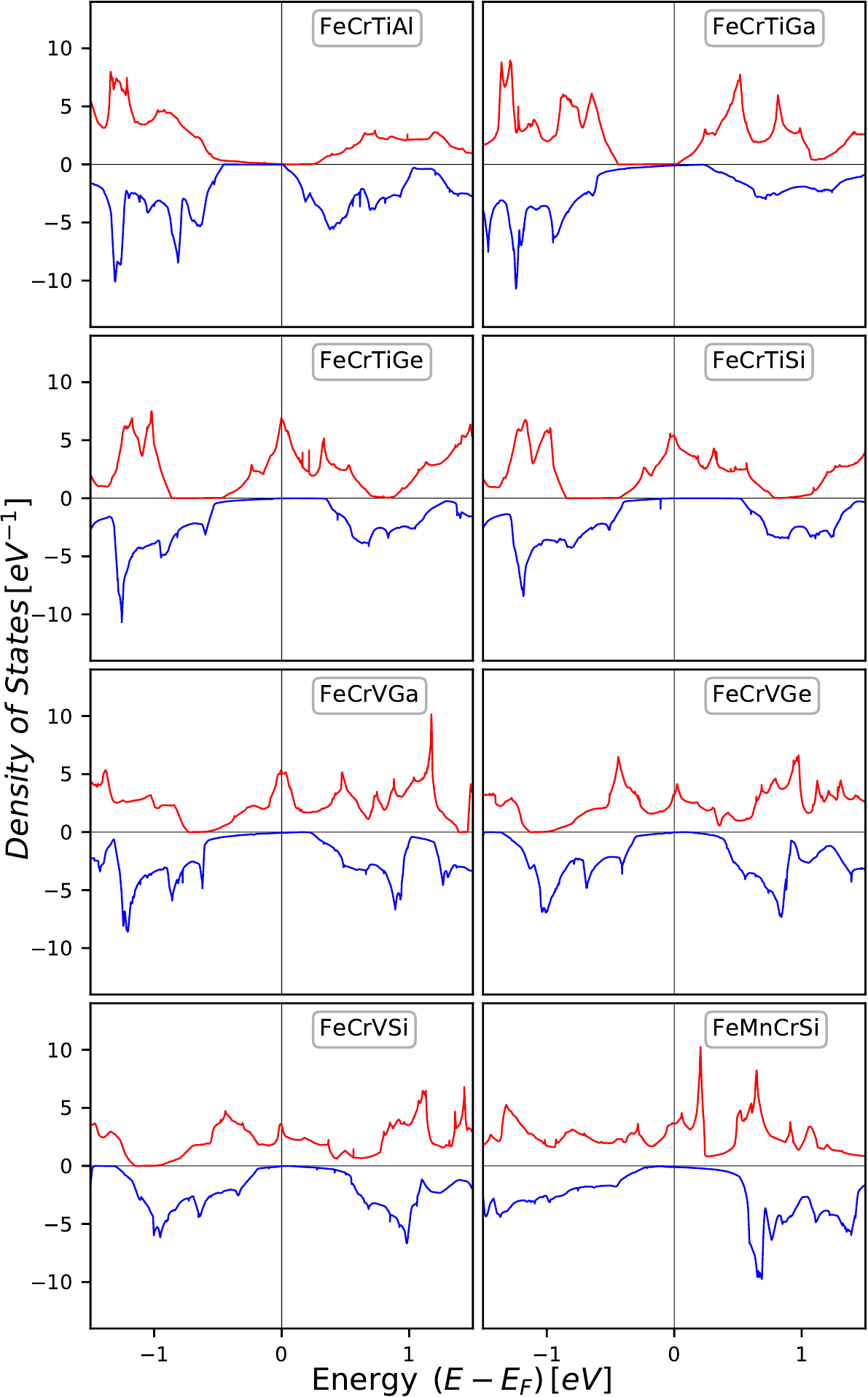}
	\caption{Spin resolved total density of states of  different quaternary Heusler alloys. Zero of x-axis correspond to Fermi level. Postive(Negative) values correspond to the spin-up(down) electrons }
	\label{fig:febased}
\end{figure}

The Slater-Pauling rule and the hybridization scheme for the Heusler alloys, developed by Galanakis \textit{et al.}\cite{Galanakis2002,Galanakisinverse,ozdougan2013slater}, is widely used to interpret the spin resolved electronic band structure of the Heusler compounds. In our past work, we have exploited this hybridization scheme to explain the observed spin resolved electronic bandstructure and magnetic moments of the inverse and full Heusler compounds \cite{our,our2}. The hybridization scheme of quaternary Heusler compounds is almost similar to full-Heusler compounds with
slight modifications in the latter case. For the quaternary Heusler compounds under investigation,the two transition metal atoms X and X\textquotesingle\, form one cube occupying the A and C co-ordinates whereas the transition metal atoms  Q and the main group element Z form another cubic sublattice. First of all, the \textit{d} orbitals of  X and X\textquotesingle\, atoms hybridize since the A and C co-ordinates are equivalent, which in turn hybridize with \textit{d}-orbitals of the transition metal atoms Q forming different bonding, anti-bonding and non-bonding states. It is worth noting that due to the symmetry of 216 space group, interchange in the position of X and X\textquotesingle\, atoms or the exchange of the location of  Q and Z atoms does not alter the structure of the crystal. Depending upon the relative postion of the \textit{d} orbitals of  Q atoms compared to X and X\textquotesingle\, atoms, the quaternary Heusler alloys can have a more complex Slater-Pauling rule. 
\begin{figure}[h!]
	\centering
	\includegraphics[width=1\linewidth]{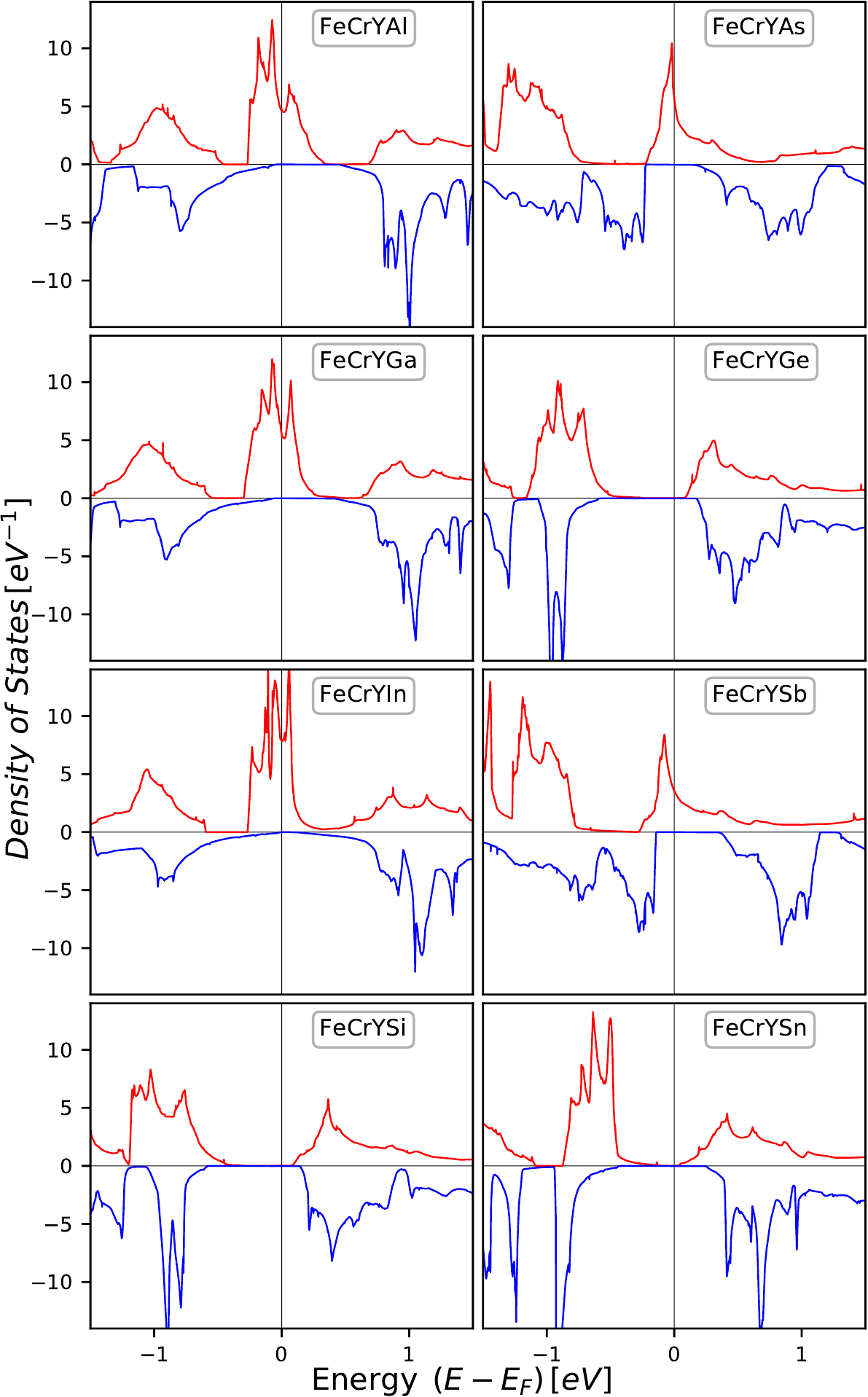}
	\caption{Spin resolved total density of states (DOS) of  different FeCrYZ [Z= Al, Ga, As, Ge, Si, Sb, In, and Sn ] quaternary Heusler alloys. Zero of x-axis correspond to Fermi level.}
	\label{fig:Ybased}
\end{figure}

From the investigation of the total spin magnetic moments of the compounds and spin-resolved bandstructure, it is clear that the Ti and V based compounds including FeMnCrSi follow the Slater-Pauling rule which can be written as
\begin{equation}
m=N_v-24
\end{equation} 
Here, $m$ is the total spin magnetic moment in $\mu_B$ per unit cell and $N_v$ is the number of valence electrons of the particular compound. Here it is important to note that for the compounds which have a number of valence electrons less than 24 and which follow the Slater-Pauling rule $m=N_v-24$, the gap remains in the spin-down band which always accommodates 12 electrons but now the spin-down electrons are the majority ones and the spin-up electrons are the minority electrons. The 12 states in the spin-down band below the Fermi level are comprised of different states which arise from main group elements and transition metals. 
\begin{figure}[h!]
	\centering
	\includegraphics[width=1\linewidth]{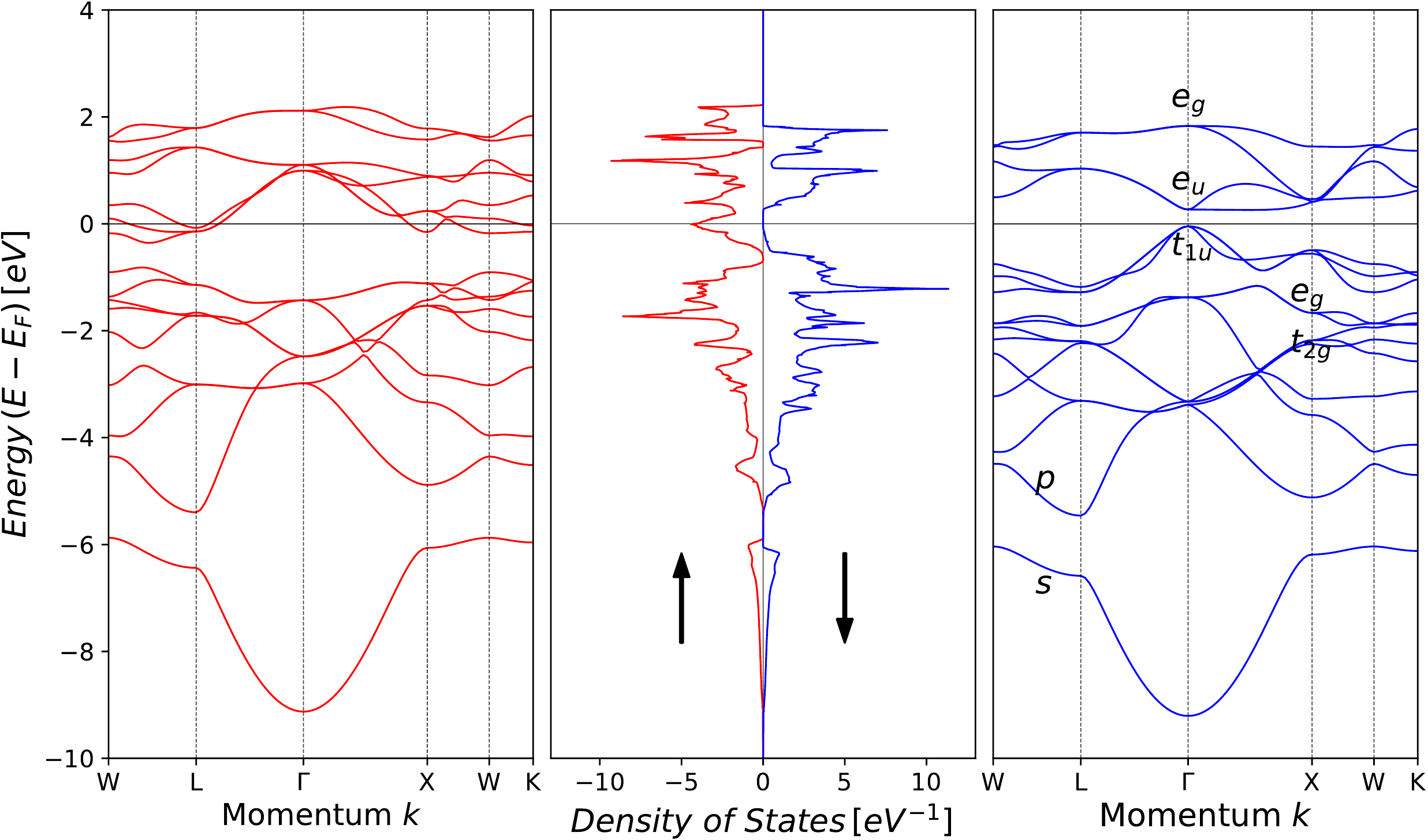}
	\caption{Spin resolved electronic bandstructure and total density of state of FeCrVAl.}
	\label{fig:fecrval}
\end{figure}

The low-lying four \textit{sp} states, one \textit{s} and three \textit{p} states, are accompanied by higher-lying eight states from the hybridization of the \textit{d} orbitals of the  transition metals. Among these eight states, uppermost triple   $t_{1u}$ states, just below the Fermi level, are followed by triplet $t_{2g}$ and double $e_g$ states as shown in the figure\ref{fig:fecrval}. The doubly degenerated bands above the Fermi level are $e_u$ and $e_g$ states respectively. Because of this, the gap in these compounds is always between non-bonding $t_{1u}$ and $e_u$ states. These states are called non-bonding as they are the states which arise from the hybridization of X and X\textquotesingle\, atoms only and hence do not have contribution from Q atoms. Since the X and X\textquotesingle\, atoms are second nearest neighbors, the hybridization between the \textit{d} orbitals of these atoms is expected to be weak which results in the small bandgap in the quaternary Heusler contrasted to half-Heusler compounds, where the bandgap is significantly larger.     Since FeMnCrSi has 25 valence electrons, which is greater than 24, for this particular case, the spin-up band is the majority band and spin-down is the minority band. The state counting and hybridization scheme remains the same as described above for this compound.  

For the Fe-Y compounds, the Slater-Pauling rule can be written as,
\begin{equation}
m=N_v-18
\end{equation} 
Here, the symbols have the usual meaning. In the Fe-Y compounds, the presence of Yttrium makes the $t_{1u}$ states to
be above the Fermi level and the gap in the spin-down band is now between
the $t_{2g}$ and the $t_{1u}$ states instead of being between the $t_{1u}$ and $e_u$
states. This is why there are now 9 spin down occupied states and it
follows the $m=N_v-18$ rule instead of the $m=N_v-24$ rule. Now the spin down are the minority spin band and the spin up is the majority spin band. The extra electrons populate the spin-up (majority) states. 
\subsection{Electronic properties} 
In table \ref{tab:Table-II}, We present the total spin magnetic moments of the compounds for three different configurations. For all the compounds, the type-III configuration shows integer (or nearly integer) value of spin magnetic moment which is the necessary condition for the half-metallicity. In figures \ref{fig:febased} and \ref{fig:Ybased}, we present the total density of states (DOS) of the compounds for the stable configuration. For every compound, we have also closely investigated the partial density of states and bandstructure which readers can  find in the supplementary information. It is interesting to analyze the DOS around the Fermi level as it reveals the exciting properties like SGS and half-metallicity. When the gap is very small around the Fermi level, it is not always easy to determine whether the compound is half-metallic or nearly half-metallic. Since the Brillouin zone is prepared by selecting a specific number of points, one should be more careful with the compounds having a small gap and hence the analysis of the total DOS, the bandstructure, and the total spin magnetic moment is necessary to list the compounds in a specific category. 
\begin{figure*}%
	\centering
	\subfloat[\centering Type-II]{{\includegraphics[width=0.49\linewidth]{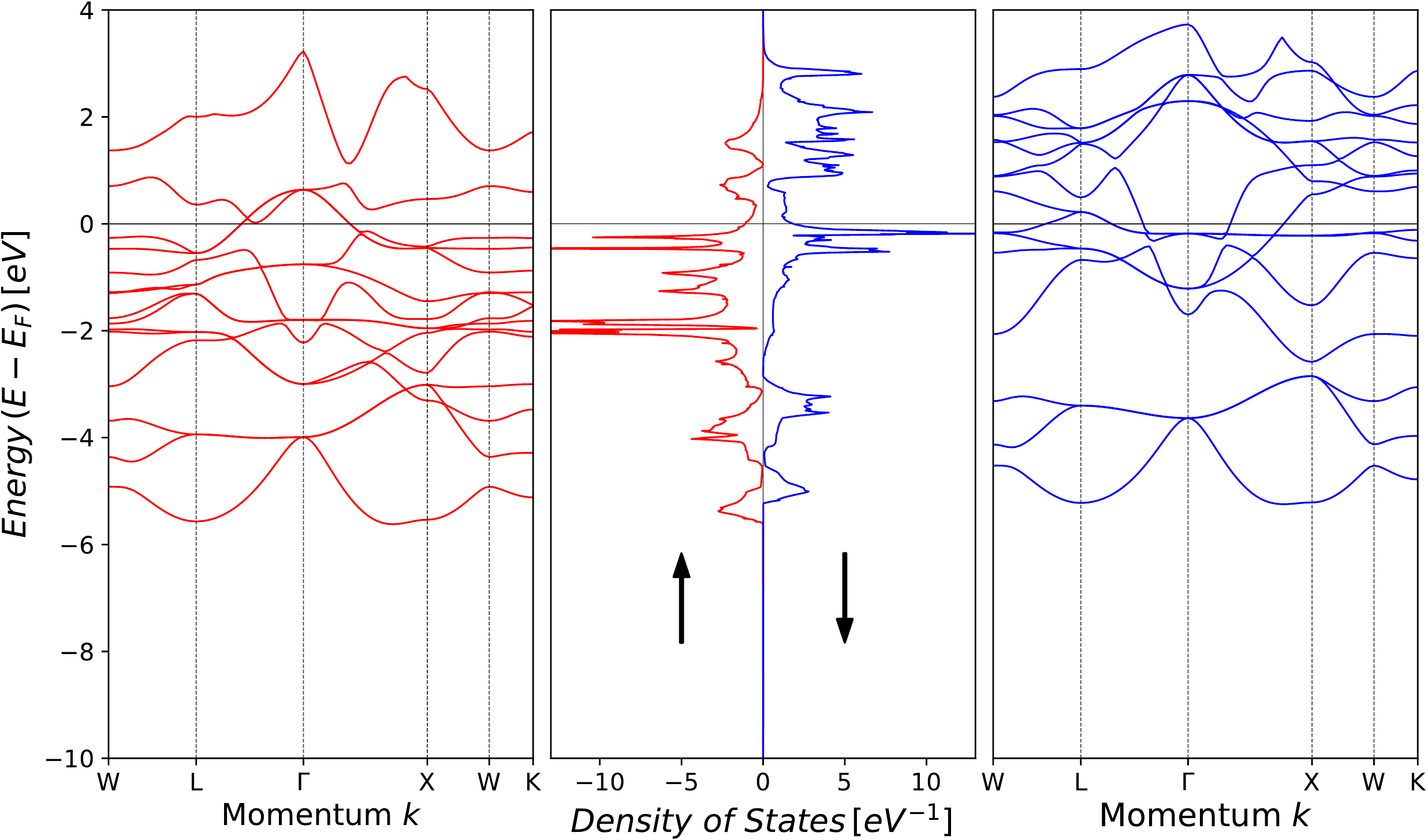} }}%
	\enspace
	\subfloat[\centering Type-III]{{\includegraphics[width=0.49\linewidth]{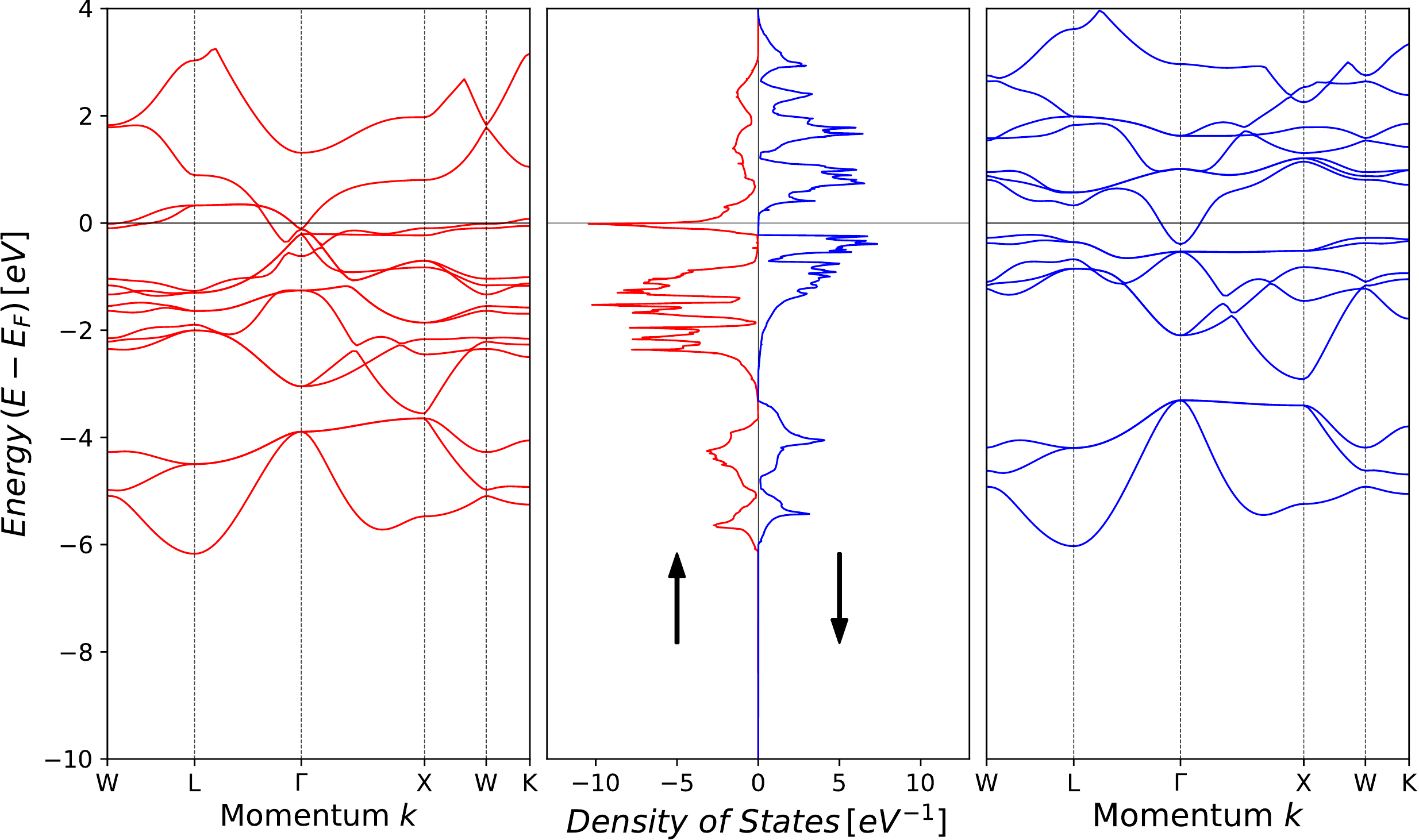} }}%
	\caption{Spin resolved electronic bandstructure and total density of states (DOS) of FeCrYAs for Type-II and Type-III configuration.}%
	\label{fig:example}%
\end{figure*}

FeCrVZ (Z= Ga, Ge, Si) and FeMnCrSi are categorized as half-metals where there is the usual metallic behavior of the spin-up channel and a half-metallic gap with minimal energy in the spin-down channel i.e. the Fermi level touches (or slightly crosses) the valence or the conduction band. FeCrVAl and FeCrTiSi are robust half-metallic compounds with a significant gap in the spin-down channel. From the meticulous study of spin-resolved DOS and bandstructure, FeCrTiGa and FeCrTiAl can be listed as type-II SGS. The perfect type-II SGS should have a gap in both spin-up and spin-down bandstructure where the conduction band minimum of the spin-up channel touches the valence band maximum of the spin-down channel at the Fermi-level. In our study, the bandstructure of both the compounds slightly deviates from the ideal type-II SGS behavior. Although there is a significant gap between the valence band maximum and the conduction band minimum of the two different spin channels, the other two alternative spin bands do not touch each other exactly at the Fermi-level preventing them from being perfect type-II SGS. This observed deviation is common in the case of type-II SGS materials \cite{aull2019ab} and the position of the Fermi-level can be tuned  within the spin-up or spin-down bandstructure because of which these compounds would be ideal candidates for re-configurable spintronics devices.

\subsection{Results on the Fe-Y compounds}

\begin{figure}[h!]
	\centering
	\includegraphics[width=1\linewidth]{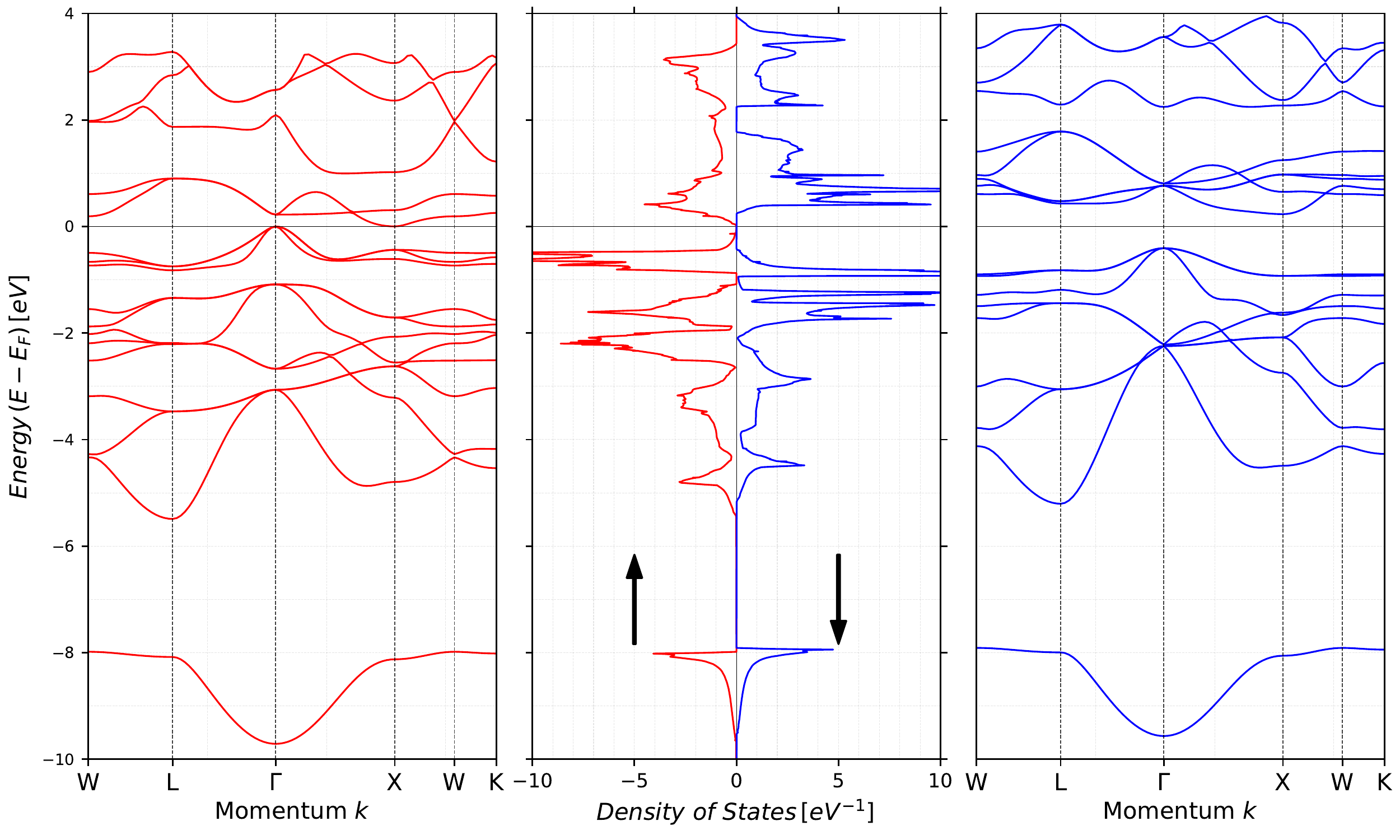}
	\caption{Spin resolved electronic bandstructure and total density of state of FeCrYSn.}
	\label{fig:FYCSn_spingapless}
\end{figure}

The Fe-Y compounds are studied to resolve the differences observed in the literature\cite{rasul2019study,idrissi2020half} and to determine the electronic and magnetic properties of some compounds. From the spin-resolved DOS presented in figure \ref{fig:Ybased}, it is clear that FeCrYZ (Z=Al, Ga, In, Sb) compounds exhibit half-metallic character as there is a gap in the minority spin channel and typical metallic character can be observed in the spin-up channel.  While the gap is very small in the case of FeCrYIn, the other three compounds possess a significant gap in the spin-down channel. FeCrYSi and FeCrYGe have a large gap in both spin directions, and the gap is situated in comparable energy region because of which they can be listed as magnetic semiconductors. From the spin-resolved bandstructure and total DOS of FeCrYSn \ref{fig:FYCSn_spingapless}, it is obvious that FeCrYSn is a type-I spin-gapless semiconductor.

The $t_{1u}$ and $e_u$ states touch the Fermi level at $\Gamma$  and $X$ point respectively in the spin-up channel whereas there is a usual $t_{2g}-t_{1u}$ gap in the spin-down channel. The close inspection of DOS around the Fermi level of FeCrYSn shows a valley approaching zero peak on the spin-up region whereas a large energy gap can be seen in the spin-down region corroborating the SGS nature of the compound.

From table \ref{tab:Table-II} it is evident that for FeCrYAs type-II structure is more stable than type-III. Since it is unusual for the quaternary Heusler alloys to have this particular structure stable, we investigated the bandstructure, partial DOS (PDOS), and total DOS for all three structures in detail for this particular compound. Authors in reference \cite{rasul2019study} have not considered the different structure for the study of Fe-Y quaternary Heusler alloys so they have assumed type-III configuration as the stable structure. We are able to reproduce the spin-resolved bandstructure and DOS of FeCrYAs, which is presented in figure\ref{fig:example} along with the DOS and bandstructure of type-II configuration. From the total DOS of type-III configuration, it is quite clear that the Fermi level is pinned exactly at the maximum of a
majority-spin peak. This type of electronic configurations can not be stable
since there is too much electronic charge at the Fermi level with the
maximum possible energy. It is exactly the same mechanism as in bulk Fe,
Co, etc. Thus the system prefers type-II configuration where the band energy is smaller  than that of type-I and type-III cases since there is less electronic charge at the
Fermi level with the maximum energy. Thus, although usually half-metallic
states correspond to the lowest energy since there is no minority-spin
DOS at the Fermi level, the extremely large majority-spin DOS at the
Fermi level makes the type-III arrangement unstable and we end up with
the type-II structure as the ground state.

In this study, we have also tried to rectify the anomalies observed in the spin magnetic moments of different Fe-Y compounds in literature\cite{rasul2019study,idrissi2020half}. For example, the calculated magnetic moment of FeCrYAl in the reference \cite{rasul2019study} is 3.89 $\mu_B$ whereas in referene \cite{idrissi2020half} the value is 2.00 $\mu_B$ for the same compound. Similarly, the total spin magnetic moment of FeCrYSb in reference \cite{idrissi2020half} is 2.30 $\mu_B$  but in reference \cite{rasul2019study} the calculated value is 4.00 $\mu_B$. As can be seen from the table, our calculated values of total spin magnetic moments for the compounds FeCrYZ (Z= Al, Sb) are 2 $\mu_B$ and 4 $\mu_B$ respectively. Hence, our computed value of the total spin magnetic moments are in agreement with \cite{idrissi2020half} for FeCrYAl but for FeCrYSb our value tends to align with  reference \cite{rasul2019study}. This is just an example, there are several Fe-Y compounds whose magnetic moments demonstrate this type of incongruous behavior which we have attempted to resolve in our calculation.   

One should always be careful while dealing with magnetic materials as the ground state of magnetic compounds is more complex to calculate than for the usual metals. In the case of magnetic materials, multiple local minima of total energy can exist. Thus, it is usual to converge to different electronic and magnetic states depending upon the initial magnetic configurations of individual atoms. Contingent on the initial configuration, the system is trapped near different local
minimum and the calculation is converged to this minimum of the
total energy. When this is the case, one should calculate the total energy for different initial configurations \textit{i.e.} ferromagnetic and antiferromagnetic configurations, and compare the converged total energies to decide which one is the ground state. Authors in reference\cite{rasul2019study} have studied the total energy  variation with respect to the cell volume for the different compounds in paramagnetic and ferromagnetic initial configurations but they have not included the case of anti-ferromagnetic phases in their study. It seems that authors in reference \cite{idrissi2020half} have assumed the initial configuration of antiferromagnetic phases for their calculations. We can assert this assumption because our calculated value of total spin 
magnetic moments for antiferromagnetic configuration, though this configuration does not result in the ground state for the compound, is 2.16 $\mu_B$, nearly equal to the value produced by the authors in the paper. Similarly, our calculated value of the total spin magnetic moments in the case of FeCrYAl for ferromagnetic phases is 3.87 $\mu_B$; the value presented by the authors in reference\cite{rasul2019study} for the compound is 3.89 $\mu_B$. Thus it can be concluded that the discrepancy observed in the spin magnetic moments of several compounds reported by the authors in references \cite{rasul2019study, idrissi2020half} is due to the faulty assumption of the ground state. 
\subsection{Magnetic properties} 

In table 3, we have gathered the total and atomic spin magnetic moments of the compounds. Except for FeCrYAs, other compounds under investigation support the half-metallic or SGS nature of the compounds as discussed above since their total spin magnetic moment is an integer (or nearly integer). It should be noted that for these compounds the value of the total spin magnetic moment is considerably large ranging from 1 to 4 $\mu_B$. The observed high spin-magnetic moment of the compounds can be attributed to the significant individual spin-magnetic moments of the constituent transition metals atoms. It is evident from the table that the contribution of Cr atoms is substantial for the observed total spin magnetic moments of the compounds in all cases. The ferromagnetic and anti-ferromagnetic alignment of the nearest neighboring (NN) atoms are different depending upon the lattice parameter and transition metal atoms involved in the compounds. For example, for the first eight compounds in table 3, both Fe and Cr are anti-ferromagnetically coupled with the NN atoms V or Ti. Similarly, for Fe-Y compounds, one can see that Cr atoms are ferromagnetically coupled with Y atoms whereas the coupling is anti-ferromagnetic for Fe and Y atoms. These different coupling of the atomic spin magnetic moment of NN atoms is dictated by the Bethe-Slater curve\citealp{jiles2015introduction} which exploits the Heisenberg model of magnetism to explain the observed anti-ferromagnetism and ferromagnetism in the 3\textit{d} transition metal atoms. The idea is that the exchange constants depend on the overlap of the 3\textit{d} wavefunctions of the neighboring atoms. In the case of the early transition metal atoms, the 3\textit{d} wavefunctions are much more extended (larger \textit{r} values) than in the case of the late transition metal atoms (small \textit{r} values). Thus given the atomic radius \textit{r} and inter-atomic distance \textit{a}, for the early ones the $a/r$ ratio is small and we have antiferromagnetism and for the late, the $a/r$ ratio is large, leading
to ferromagnetism. If we increase the interatomic distance \textit{a} between two NN early transition metal atoms then the $a/r$ ratio becomes comparable to the Fe case and we get a transition from antiferromagnetism to ferromagnetism.

For the first eight compounds the overlap between the 3\textit{d }electrons of the
nearest neighboring  Cr and V atoms are comparable to bulk Cr leading to the
antiferromagnetic coupling of their spin magnetic moments since they are
early transition metal atoms. V atoms mediate the interactions between
the Fe and the Cr atoms. The distance between the neighboring V and Fe
atoms is also quite small and the overlap between the 3\textit{d} orbitals of the
V and the Fe atoms is  sizeable leading to the antiferromagnetic coupling
between the Fe and the V spin moments. The same argument is true when the mediating atoms is Ti.

In the case of Fe-Y compounds, the lattice constants are larger and thus the distance between Cr and Y becomes larger leading to ferromagnetic
coupling. One could expect that also the Fe-Y spin moment should be
ferromagnetic due to the larger interatomic distance but the overlap scheme is much more complex since there is also the Fe-Cr overlap leading finally to negative Fe spin moments. For FeCrYAs and FeCrYSb compounds, the complexity of the interactions leads to different behavior and the explanation is not always straightforward.
\section{Conclusion}
In summary, we have presented detailed \textit{ab intio} calculations of FeCr-based quaternary Heusler alloys. The electronic and magnetic properties are studied along with different mechanical parameters which are important to understand the nature of the compounds. The thermodynamic stability is ensured by studying the convex-hull distance whereas mechanical and dynamical stability of the compounds is confirmed by calculating the elastic constants and phonon dispersion curves. The interesting properties like half-metallicity and spin-gapless semiconducting behavior is observed in the studied compounds. In addition, a comprehensive study of the Fe-Y compounds is carried out to resolve the contradicting results observed in the literature. We believe that our study will trigger a further exploration of FeCr-based quaternary Heusler alloys.                      
\section{Acknowledgement}
Part of the calculation was performed with the computational resources provided by the Kathmandu University Supercomputer Center established with the equipment donated by CERN.

\bibliography{mybibfile}

\begin{thebibliography}{55}
\expandafter\ifx\csname natexlab\endcsname\relax\def\natexlab#1{#1}\fi
\providecommand{\url}[1]{\texttt{#1}}
\providecommand{\href}[2]{#2}
\providecommand{\path}[1]{#1}
\providecommand{\DOIprefix}{doi:}
\providecommand{\ArXivprefix}{arXiv:}
\providecommand{\URLprefix}{URL: }
\providecommand{\Pubmedprefix}{pmid:}
\providecommand{\doi}[1]{\href{http://dx.doi.org/#1}{\path{#1}}}
\providecommand{\Pubmed}[1]{\href{pmid:#1}{\path{#1}}}
\providecommand{\bibinfo}[2]{#2}
\ifx\xfnm\relax \def\xfnm[#1]{\unskip,\space#1}\fi
\bibitem[{\ifmmode \check{Z}\else \v{Z}\fi{}uti\ifmmode~\acute{c}\else
  \'{c}\fi{} et~al.(2004)\ifmmode \check{Z}\else
  \v{Z}\fi{}uti\ifmmode~\acute{c}\else \'{c}\fi{}, Fabian, and
  Das~Sarma}]{spintronics}
\bibinfo{author}{I.~\ifmmode \check{Z}\else
  \v{Z}\fi{}uti\ifmmode~\acute{c}\else \'{c}\fi{}},
  \bibinfo{author}{J.~Fabian}, \bibinfo{author}{S.~Das~Sarma},
\newblock \bibinfo{title}{Spintronics: Fundamentals and applications},
\newblock \bibinfo{journal}{Rev. Mod. Phys.} \bibinfo{volume}{76}
  (\bibinfo{year}{2004}) \bibinfo{pages}{323--410}.
\bibitem[{Hirohata and Takanashi(2014)}]{Hirohata}
\bibinfo{author}{A.~Hirohata}, \bibinfo{author}{K.~Takanashi},
\newblock \bibinfo{title}{Future perspectives for spintronic devices},
\newblock \bibinfo{journal}{Journal of Physics D: Applied Physics}
  \bibinfo{volume}{47} (\bibinfo{year}{2014}) \bibinfo{pages}{193001}.
\bibitem[{Fong et~al.(2013)Fong, Pask, and Yang}]{fong2013half}
\bibinfo{author}{C.-y. Fong}, \bibinfo{author}{J.~E. Pask},
  \bibinfo{author}{L.~H. Yang}, \bibinfo{title}{Half-metallic materials and
  their properties}, volume~\bibinfo{volume}{2}, \bibinfo{publisher}{World
  Scientific}, \bibinfo{year}{2013}.
\bibitem[{Felser and Hirohata(2005)}]{HeuslerPropandGrowth}
\bibinfo{editor}{C.~Felser}, \bibinfo{editor}{A.~Hirohata} (Eds.),
  \bibinfo{title}{Heusler Alloys: Properties, Growth, Applications}, volume
  \bibinfo{volume}{222} of \textit{\bibinfo{series}{Springer Series in
  Materials Science}}, \bibinfo{edition}{1} ed., \bibinfo{publisher}{Springer
  International Publishing}, \bibinfo{address}{Switzerland},
  \bibinfo{year}{2005}.
\bibitem[{Felser and Fecher(2013)}]{SpintronicsClaudia}
\bibinfo{editor}{C.~Felser}, \bibinfo{editor}{G.~H. Fecher} (Eds.),
  \bibinfo{title}{Spintronics: From Materials to Devices}, \bibinfo{edition}{1}
  ed., \bibinfo{publisher}{Springer}, \bibinfo{address}{Netherlands},
  \bibinfo{year}{2013}.
\bibitem[{Galanakis and Dederichs(2005)}]{HMAlloys-Galanakis-lec}
\bibinfo{editor}{I.~Galanakis}, \bibinfo{editor}{P.~H. Dederichs} (Eds.),
  \bibinfo{title}{Half-metallic Alloys: Fundamentals and Applications}, volume
  \bibinfo{volume}{676} of \textit{\bibinfo{series}{Lecture Notes in Physics}},
  \bibinfo{edition}{1} ed., \bibinfo{publisher}{Springer-Verlag},
  \bibinfo{address}{Berlin Heidelberg}, \bibinfo{year}{2005}.
\bibitem[{Galanakis et~al.(2002)Galanakis, Dederichs, and
  Papanikolaou}]{Galanakis2002}
\bibinfo{author}{I.~Galanakis}, \bibinfo{author}{P.~H. Dederichs},
  \bibinfo{author}{N.~Papanikolaou},
\newblock \bibinfo{title}{Slater-pauling behavior and origin of the
  half-metallicity of the full-{Heusler} alloys},
\newblock \bibinfo{journal}{Phys. Rev. B} \bibinfo{volume}{66}
  (\bibinfo{year}{2002}) \bibinfo{pages}{174429}.
\bibitem[{Skaftouros et~al.(2013)Skaftouros, Özdoğan, Şaşıoğlu, and
  Galanakis}]{Galanakisinverse}
\bibinfo{author}{S.~Skaftouros}, \bibinfo{author}{K.~Özdoğan},
  \bibinfo{author}{E.~Şaşıoğlu}, \bibinfo{author}{I.~Galanakis},
\newblock \bibinfo{title}{Generalized {Slater-Pauling} rule for the inverse
  {Heusler} compounds},
\newblock \bibinfo{journal}{Phys. Rev. B} \bibinfo{volume}{87}
  (\bibinfo{year}{2013}) \bibinfo{pages}{024420}.
\bibitem[{Kandpal et~al.(2007)Kandpal, Fecher, and Felser}]{Kandpalmainpaper}
\bibinfo{author}{H.~C. Kandpal}, \bibinfo{author}{G.~H. Fecher},
  \bibinfo{author}{C.~Felser},
\newblock \bibinfo{title}{Calculated electronic and magnetic properties of the
  half-metallic, transition metal based {Heusler} compounds},
\newblock \bibinfo{journal}{Journal of Physics D: Applied Physics}
  \bibinfo{volume}{40} (\bibinfo{year}{2007}) \bibinfo{pages}{1507--1523}.
\bibitem[{Dai et~al.(2009)Dai, Liu, Fecher, Felser, Li, and Liu}]{dai2009new}
\bibinfo{author}{X.~Dai}, \bibinfo{author}{G.~Liu}, \bibinfo{author}{G.~H.
  Fecher}, \bibinfo{author}{C.~Felser}, \bibinfo{author}{Y.~Li},
  \bibinfo{author}{H.~Liu},
\newblock \bibinfo{title}{New quarternary half metallic material {CoFeMnSi}},
\newblock \bibinfo{journal}{Journal of Applied Physics} \bibinfo{volume}{105}
  (\bibinfo{year}{2009}) \bibinfo{pages}{07E901}.
\bibitem[{Alijani et~al.(2011{\natexlab{a}})Alijani, Winterlik, Fecher,
  Naghavi, and Felser}]{alijani2011quaternary}
\bibinfo{author}{V.~Alijani}, \bibinfo{author}{J.~Winterlik},
  \bibinfo{author}{G.~H. Fecher}, \bibinfo{author}{S.~S. Naghavi},
  \bibinfo{author}{C.~Felser},
\newblock \bibinfo{title}{Quaternary half-metallic {Heusler} ferromagnets for
  spintronics applications},
\newblock \bibinfo{journal}{Physical Review B} \bibinfo{volume}{83}
  (\bibinfo{year}{2011}{\natexlab{a}}) \bibinfo{pages}{184428}.
\bibitem[{Alijani et~al.(2011{\natexlab{b}})Alijani, Ouardi, Fecher, Winterlik,
  Naghavi, Kozina, Stryganyuk, Felser, Ikenaga, Yamashita
  et~al.}]{alijani2011electronic}
\bibinfo{author}{V.~Alijani}, \bibinfo{author}{S.~Ouardi},
  \bibinfo{author}{G.~H. Fecher}, \bibinfo{author}{J.~Winterlik},
  \bibinfo{author}{S.~S. Naghavi}, \bibinfo{author}{X.~Kozina},
  \bibinfo{author}{G.~Stryganyuk}, \bibinfo{author}{C.~Felser},
  \bibinfo{author}{E.~Ikenaga}, \bibinfo{author}{Y.~Yamashita}, et~al.,
\newblock \bibinfo{title}{Electronic, structural, and magnetic properties of
  the half-metallic ferromagnetic quaternary {Heusler} compounds {CoFeMn Z (Z=
  Al, Ga, Si, Ge)}},
\newblock \bibinfo{journal}{Physical Review B} \bibinfo{volume}{84}
  (\bibinfo{year}{2011}{\natexlab{b}}) \bibinfo{pages}{224416}.
\bibitem[{Xu et~al.(2013)Xu, Liu, Du, Li, Liu, Wang, and Wu}]{xu2013new}
\bibinfo{author}{G.~Xu}, \bibinfo{author}{E.~Liu}, \bibinfo{author}{Y.~Du},
  \bibinfo{author}{G.~Li}, \bibinfo{author}{G.~Liu}, \bibinfo{author}{W.~Wang},
  \bibinfo{author}{G.~Wu},
\newblock \bibinfo{title}{A new spin gapless semiconductors family: {Quaternary
  Heusler} compounds},
\newblock \bibinfo{journal}{EPL (Europhysics Letters)} \bibinfo{volume}{102}
  (\bibinfo{year}{2013}) \bibinfo{pages}{17007}.
\bibitem[{Bainsla et~al.(2015)Bainsla, Mallick, Raja, Nigam, Varaprasad,
  Takahashi, Alam, Suresh, and Hono}]{bainsla2015spin}
\bibinfo{author}{L.~Bainsla}, \bibinfo{author}{A.~Mallick},
  \bibinfo{author}{M.~M. Raja}, \bibinfo{author}{A.~Nigam},
  \bibinfo{author}{B.~C.~S. Varaprasad}, \bibinfo{author}{Y.~Takahashi},
  \bibinfo{author}{A.~Alam}, \bibinfo{author}{K.~Suresh},
  \bibinfo{author}{K.~Hono},
\newblock \bibinfo{title}{Spin gapless semiconducting behavior in equiatomic
  quaternary {CoFeMnSi} {Heusler} alloy},
\newblock \bibinfo{journal}{Physical Review B} \bibinfo{volume}{91}
  (\bibinfo{year}{2015}) \bibinfo{pages}{104408}.
\bibitem[{Gao et~al.(2013)Gao, Hu, Yao, Luo, and Liu}]{gao2013large}
\bibinfo{author}{G.~Gao}, \bibinfo{author}{L.~Hu}, \bibinfo{author}{K.~Yao},
  \bibinfo{author}{B.~Luo}, \bibinfo{author}{N.~Liu},
\newblock \bibinfo{title}{Large half-metallic gaps in the quaternary {Heusler}
  alloys {CoFeCrZ (Z= Al, Si, Ga, Ge)}: A first-principles study},
\newblock \bibinfo{journal}{Journal of alloys and compounds}
  \bibinfo{volume}{551} (\bibinfo{year}{2013}) \bibinfo{pages}{539--543}.
\bibitem[{{\"O}zdo{\u{g}}an et~al.(2013){\"O}zdo{\u{g}}an,
  {\c{S}}a{\c{s}}{\i}o{\u{g}}lu, and Galanakis}]{ozdougan2013slater}
\bibinfo{author}{K.~{\"O}zdo{\u{g}}an},
  \bibinfo{author}{E.~{\c{S}}a{\c{s}}{\i}o{\u{g}}lu},
  \bibinfo{author}{I.~Galanakis},
\newblock \bibinfo{title}{Slater-pauling behavior in {LiMgPdSn}-type
  multifunctional quaternary {Heusler} materials: {Half-metallicity},
  spin-gapless and magnetic semiconductors},
\newblock \bibinfo{journal}{Journal of Applied Physics} \bibinfo{volume}{113}
  (\bibinfo{year}{2013}) \bibinfo{pages}{193903}.
\bibitem[{Bainsla et~al.(2015{\natexlab{a}})Bainsla, Mallick, Coelho, Nigam,
  Varaprasad, Takahashi, Alam, Suresh, and Hono}]{BAINSLA201582}
\bibinfo{author}{L.~Bainsla}, \bibinfo{author}{A.~Mallick},
  \bibinfo{author}{A.~Coelho}, \bibinfo{author}{A.~Nigam},
  \bibinfo{author}{B.~Varaprasad}, \bibinfo{author}{Y.~Takahashi},
  \bibinfo{author}{A.~Alam}, \bibinfo{author}{K.~Suresh},
  \bibinfo{author}{K.~Hono},
\newblock \bibinfo{title}{High spin polarization and spin splitting in
  equiatomic quaternary {CoFeCrAl} {Heusler} alloy},
\newblock \bibinfo{journal}{Journal of Magnetism and Magnetic Materials}
  \bibinfo{volume}{394} (\bibinfo{year}{2015}{\natexlab{a}}) \bibinfo{pages}{82
  -- 86}.
\bibitem[{Bainsla et~al.(2015{\natexlab{b}})Bainsla, Mallick, Raja, Coelho,
  Nigam, Johnson, Alam, and Suresh}]{bainsla2015origin}
\bibinfo{author}{L.~Bainsla}, \bibinfo{author}{A.~Mallick},
  \bibinfo{author}{M.~M. Raja}, \bibinfo{author}{A.~Coelho},
  \bibinfo{author}{A.~Nigam}, \bibinfo{author}{D.~D. Johnson},
  \bibinfo{author}{A.~Alam}, \bibinfo{author}{K.~Suresh},
\newblock \bibinfo{title}{Origin of spin gapless semiconductor behavior in
  {CoFeCrGa: Theory and Experiment}},
\newblock \bibinfo{journal}{Physical Review B} \bibinfo{volume}{92}
  (\bibinfo{year}{2015}{\natexlab{b}}) \bibinfo{pages}{045201}.
\bibitem[{Gao et~al.(2019)Gao, Opahle, and Zhang}]{gao2019high}
\bibinfo{author}{Q.~Gao}, \bibinfo{author}{I.~Opahle},
  \bibinfo{author}{H.~Zhang},
\newblock \bibinfo{title}{High-throughput screening for spin-gapless
  semiconductors in quaternary {Heusler} compounds},
\newblock \bibinfo{journal}{Physical Review Materials} \bibinfo{volume}{3}
  (\bibinfo{year}{2019}) \bibinfo{pages}{024410}.
\bibitem[{Aull et~al.(2019)Aull, {\c{S}}a{\c{s}}{\i}o{\u{g}}lu, Maznichenko,
  Ostanin, Ernst, Mertig, and Galanakis}]{aull2019ab}
\bibinfo{author}{T.~Aull}, \bibinfo{author}{E.~{\c{S}}a{\c{s}}{\i}o{\u{g}}lu},
  \bibinfo{author}{I.~Maznichenko}, \bibinfo{author}{S.~Ostanin},
  \bibinfo{author}{A.~Ernst}, \bibinfo{author}{I.~Mertig},
  \bibinfo{author}{I.~Galanakis},
\newblock \bibinfo{title}{Ab initio design of quaternary {Heusler} compounds
  for reconfigurable magnetic tunnel diodes and transistors},
\newblock \bibinfo{journal}{Physical Review Materials} \bibinfo{volume}{3}
  (\bibinfo{year}{2019}) \bibinfo{pages}{124415}.
\bibitem[{Guo et~al.(2018)Guo, Ni, Liang, and Luo}]{guo2018magnetic}
\bibinfo{author}{X.~Guo}, \bibinfo{author}{Z.~Ni}, \bibinfo{author}{Z.~Liang},
  \bibinfo{author}{H.~Luo},
\newblock \bibinfo{title}{Magnetic semiconductors and half-metals in
  {FeRu}-based quaternary {Heusler} alloys},
\newblock \bibinfo{journal}{Computational Materials Science}
  \bibinfo{volume}{154} (\bibinfo{year}{2018}) \bibinfo{pages}{442--448}.
\bibitem[{Luo et~al.(2020)Luo, Li, Sun, Liu, and Liang}]{siteRu}
\bibinfo{author}{H.~Luo}, \bibinfo{author}{Q.~Li}, \bibinfo{author}{K.~Sun},
  \bibinfo{author}{S.~Liu}, \bibinfo{author}{Z.~Liang},
\newblock \bibinfo{title}{Magnetic properties and site preference of {Ru} in
  heusler alloys {Fe$_2$V$_{1-x}$Ru$_x$Si} (x= 0.25, 0.5, 0.75, 1)},
\newblock \bibinfo{journal}{Journal of Magnetism and Magnetic Materials}
  \bibinfo{volume}{496} (\bibinfo{year}{2020}) \bibinfo{pages}{165908}.
\bibitem[{Ray et~al.(2021)Ray, Kaphle, Rai, Yadav, Paudel, and
  Paudyal}]{ray2021strain}
\bibinfo{author}{R.~B. Ray}, \bibinfo{author}{G.~C. Kaphle},
  \bibinfo{author}{R.~K. Rai}, \bibinfo{author}{D.~K. Yadav},
  \bibinfo{author}{R.~Paudel}, \bibinfo{author}{D.~Paudyal},
\newblock \bibinfo{title}{Strain induced electronic structure, and magnetic and
  structural properties in quaternary {Heusler} alloys {ZrRhTiZ (Z= Al, In)}},
\newblock \bibinfo{journal}{Journal of Alloys and Compounds}
  \bibinfo{volume}{867} (\bibinfo{year}{2021}) \bibinfo{pages}{158906}.
\bibitem[{Rasul et~al.(2019)Rasul, Javed, Khan, and Hussain}]{rasul2019study}
\bibinfo{author}{M.~N. Rasul}, \bibinfo{author}{A.~Javed},
  \bibinfo{author}{M.~A. Khan}, \bibinfo{author}{A.~Hussain},
\newblock \bibinfo{title}{Study of the structural, mechanical, electronic and
  magnetic properties of quaternary {YFeCrX (X= Al, Ga, In, Si, Ge, Sn, P, As,
  Sb)} {Heusler} alloys},
\newblock \bibinfo{journal}{Journal of Magnetism and Magnetic Materials}
  \bibinfo{volume}{476} (\bibinfo{year}{2019}) \bibinfo{pages}{398--411}.
\bibitem[{Idrissi et~al.(2020)Idrissi, Ziti, Labrim, Bahmad, El~Housni,
  Khalladi, Mtougui, and El~Mekkaoui}]{idrissi2020half}
\bibinfo{author}{S.~Idrissi}, \bibinfo{author}{S.~Ziti},
  \bibinfo{author}{H.~Labrim}, \bibinfo{author}{L.~Bahmad},
  \bibinfo{author}{I.~El~Housni}, \bibinfo{author}{R.~Khalladi},
  \bibinfo{author}{S.~Mtougui}, \bibinfo{author}{N.~El~Mekkaoui},
\newblock \bibinfo{title}{Half-metallic behavior and magnetic proprieties of
  the quaternary {Heusler} alloys {YFeCrZ (Z= Al, Sb and Sn)}},
\newblock \bibinfo{journal}{Journal of Alloys and Compounds}
  \bibinfo{volume}{820} (\bibinfo{year}{2020}) \bibinfo{pages}{153373}.
\bibitem[{Shakil et~al.(2021)Shakil, Arshad, Aziz, Gillani, Rizwan, and
  Zafar}]{SHAKIL2021157370}
\bibinfo{author}{M.~Shakil}, \bibinfo{author}{H.~Arshad},
  \bibinfo{author}{S.~Aziz}, \bibinfo{author}{S.~Gillani},
  \bibinfo{author}{M.~Rizwan}, \bibinfo{author}{M.~Zafar},
\newblock \bibinfo{title}{Determination of phase stability, half metallicity,
  mechanical and thermal behavior of {Fe} based quaternary {Heusler} alloys},
\newblock \bibinfo{journal}{Journal of Alloys and Compounds}
  \bibinfo{volume}{856} (\bibinfo{year}{2021}) \bibinfo{pages}{157370}.
\bibitem[{Saal et~al.(2013)Saal, Kirklin, Aykol, Meredig, and
  Wolverton}]{saal2013materials}
\bibinfo{author}{J.~E. Saal}, \bibinfo{author}{S.~Kirklin},
  \bibinfo{author}{M.~Aykol}, \bibinfo{author}{B.~Meredig},
  \bibinfo{author}{C.~Wolverton},
\newblock \bibinfo{title}{Materials design and discovery with high-throughput
  density functional theory: the open quantum materials database {(OQMD)}},
\newblock \bibinfo{journal}{Jom} \bibinfo{volume}{65} (\bibinfo{year}{2013})
  \bibinfo{pages}{1501--1509}.
\bibitem[{Kirklin et~al.(2015)Kirklin, Saal, Meredig, Thompson, Doak, Aykol,
  R{\"u}hl, and Wolverton}]{OQMDkirklin2015}
\bibinfo{author}{S.~Kirklin}, \bibinfo{author}{J.~E. Saal},
  \bibinfo{author}{B.~Meredig}, \bibinfo{author}{A.~Thompson},
  \bibinfo{author}{J.~W. Doak}, \bibinfo{author}{M.~Aykol},
  \bibinfo{author}{S.~R{\"u}hl}, \bibinfo{author}{C.~Wolverton},
\newblock \bibinfo{title}{The open quantum materials database (oqmd): assessing
  the accuracy of dft formation energies},
\newblock \bibinfo{journal}{npj Computational Materials} \bibinfo{volume}{1}
  (\bibinfo{year}{2015}) \bibinfo{pages}{1--15}.
\bibitem[{Galanakis et~al.(2014)Galanakis, Özdoğan, and
  Şaşıoğlu}]{galanakis2014crvtial}
\bibinfo{author}{I.~Galanakis}, \bibinfo{author}{K.~Özdoğan},
  \bibinfo{author}{E.~Şaşıoğlu},
\newblock \bibinfo{title}{High-t c fully compensated ferrimagnetic
  semiconductors as spin-filter materials: the case of {CrVXAl (X= Ti, Zr, Hf)
  Heusler} compounds},
\newblock \bibinfo{journal}{Journal of Physics: Condensed Matter}
  \bibinfo{volume}{26} (\bibinfo{year}{2014}) \bibinfo{pages}{086003}.
\bibitem[{Venkateswara et~al.(2018)Venkateswara, Gupta, Samatham, Varma,
  Suresh, Alam et~al.}]{crvtial_exp}
\bibinfo{author}{Y.~Venkateswara}, \bibinfo{author}{S.~Gupta},
  \bibinfo{author}{S.~S. Samatham}, \bibinfo{author}{M.~R. Varma},
  \bibinfo{author}{K.~Suresh}, \bibinfo{author}{A.~Alam}, et~al.,
\newblock \bibinfo{title}{Competing magnetic and spin-gapless semiconducting
  behavior in fully compensated ferrimagnetic {CrVTiAl}: {Theory} and
  experiment},
\newblock \bibinfo{journal}{Physical Review B} \bibinfo{volume}{97}
  (\bibinfo{year}{2018}) \bibinfo{pages}{054407}.
\bibitem[{Giannozzi et~al.(2009)Giannozzi, Baroni, Bonini, Calandra, Car, and
  \textit{et al.}}]{QE-2009}
\bibinfo{author}{P.~Giannozzi}, \bibinfo{author}{S.~Baroni},
  \bibinfo{author}{N.~Bonini}, \bibinfo{author}{M.~Calandra},
  \bibinfo{author}{R.~Car}, \bibinfo{author}{\textit{et al.}},
\newblock \bibinfo{title}{{QUANTUM ESPRESSO}: a modular and open-source
  software project for quantum simulations of materials},
\newblock \bibinfo{journal}{Journal of Physics: Condensed Matter}
  \bibinfo{volume}{21} (\bibinfo{year}{2009}) \bibinfo{pages}{395502 (19pp)}.
\bibitem[{Giannozzi et~al.(2017)Giannozzi, Andreussi, Brumme, Bunau, Nardelli,
  and \textit{et al.}}]{QE-2017}
\bibinfo{author}{P.~Giannozzi}, \bibinfo{author}{O.~Andreussi},
  \bibinfo{author}{T.~Brumme}, \bibinfo{author}{O.~Bunau},
  \bibinfo{author}{M.~B. Nardelli}, \bibinfo{author}{\textit{et al.}},
\newblock \bibinfo{title}{Advanced capabilities for materials modelling with
  quantum espresso},
\newblock \bibinfo{journal}{Journal of Physics: Condensed Matter}
  \bibinfo{volume}{29} (\bibinfo{year}{2017}) \bibinfo{pages}{465901}.
\bibitem[{Perdew et~al.(1996)Perdew, Burke, and
  Ernzerhof}]{perdew1996generalized}
\bibinfo{author}{J.~P. Perdew}, \bibinfo{author}{K.~Burke},
  \bibinfo{author}{M.~Ernzerhof},
\newblock \bibinfo{title}{Generalized gradient approximation made simple},
\newblock \bibinfo{journal}{Physical review letters} \bibinfo{volume}{77}
  (\bibinfo{year}{1996}) \bibinfo{pages}{3865}.
\bibitem[{Bl\"ochl et~al.(1994)Bl\"ochl, Jepsen, and
  Andersen}]{lineartetrahedra}
\bibinfo{author}{P.~E. Bl\"ochl}, \bibinfo{author}{O.~Jepsen},
  \bibinfo{author}{O.~K. Andersen},
\newblock \bibinfo{title}{Improved tetrahedron method for brillouin-zone
  integrations},
\newblock \bibinfo{journal}{Phys. Rev. B} \bibinfo{volume}{49}
  (\bibinfo{year}{1994}) \bibinfo{pages}{16223--16233}.
\bibitem[{Togo and Tanaka(2015)}]{phonopy}
\bibinfo{author}{A.~Togo}, \bibinfo{author}{I.~Tanaka},
\newblock \bibinfo{title}{First principles phonon calculations in materials
  science},
\newblock \bibinfo{journal}{Scr. Mater.} \bibinfo{volume}{108}
  (\bibinfo{year}{2015}) \bibinfo{pages}{1--5}.
\bibitem[{Ma et~al.(2018)Ma, He, Mazumdar, Munira, Keshavarz, Lovorn,
  Wolverton, Ghosh, and Butler}]{ma2018computational}
\bibinfo{author}{J.~Ma}, \bibinfo{author}{J.~He},
  \bibinfo{author}{D.~Mazumdar}, \bibinfo{author}{K.~Munira},
  \bibinfo{author}{S.~Keshavarz}, \bibinfo{author}{T.~Lovorn},
  \bibinfo{author}{C.~Wolverton}, \bibinfo{author}{A.~W. Ghosh},
  \bibinfo{author}{W.~H. Butler},
\newblock \bibinfo{title}{Computational investigation of inverse {Heusler}
  compounds for spintronics applications},
\newblock \bibinfo{journal}{Physical Review B} \bibinfo{volume}{98}
  (\bibinfo{year}{2018}) \bibinfo{pages}{094410}.
\bibitem[{Kresse and Furthm\"uller(1996)}]{vasp}
\bibinfo{author}{G.~Kresse}, \bibinfo{author}{J.~Furthm\"uller},
\newblock \bibinfo{title}{Efficient iterative schemes for ab initio
  total-energy calculations using a plane-wave basis set},
\newblock \bibinfo{journal}{Phys. Rev. B} \bibinfo{volume}{54}
  (\bibinfo{year}{1996}) \bibinfo{pages}{11169--11186}.
\bibitem[{Kresse and Joubert(1999)}]{PAW}
\bibinfo{author}{G.~Kresse}, \bibinfo{author}{D.~Joubert},
\newblock \bibinfo{title}{From ultrasoft pseudopotentials to the projector
  augmented-wave method},
\newblock \bibinfo{journal}{Phys. Rev. B} \bibinfo{volume}{59}
  (\bibinfo{year}{1999}) \bibinfo{pages}{1758--1775}.
\bibitem[{Wu et~al.(2013)Wu, Lazic, Hautier, Persson, and Ceder}]{wu2013first}
\bibinfo{author}{Y.~Wu}, \bibinfo{author}{P.~Lazic},
  \bibinfo{author}{G.~Hautier}, \bibinfo{author}{K.~Persson},
  \bibinfo{author}{G.~Ceder},
\newblock \bibinfo{title}{First principles high throughput screening of
  oxynitrides for water-splitting photocatalysts},
\newblock \bibinfo{journal}{Energy \& environmental science}
  \bibinfo{volume}{6} (\bibinfo{year}{2013}) \bibinfo{pages}{157--168}.
\bibitem[{ics(2020)}]{icsd}
\bibinfo{title}{{Inorganic Crystal Structure Database} [{ICSD}]},
  \bibinfo{year}{2020}. \URLprefix \url{https://icsd.fiz-karlsruhe.de/}.
\bibitem[{Golesorkhtabar et~al.(2013)Golesorkhtabar, Pavone, Spitaler,
  Puschnig, and Draxl}]{elasticTool}
\bibinfo{author}{R.~Golesorkhtabar}, \bibinfo{author}{P.~Pavone},
  \bibinfo{author}{J.~Spitaler}, \bibinfo{author}{P.~Puschnig},
  \bibinfo{author}{C.~Draxl},
\newblock \bibinfo{title}{{ElaStic}: A tool for calculating second-order
  elastic constants from first principles},
\newblock \bibinfo{journal}{Computer Physics Communications}
  \bibinfo{volume}{184} (\bibinfo{year}{2013}) \bibinfo{pages}{1861 -- 1873}.
\bibitem[{Born and Huang(1956)}]{born_huang_1956}
\bibinfo{author}{M.~Born}, \bibinfo{author}{K.~Huang},
  \bibinfo{title}{Dynamical theory of crystal lattices},
  \bibinfo{publisher}{Clarendon Press}, \bibinfo{year}{1956}.
\bibitem[{Bhagavantam and Bhimasenachar(1944)}]{diamond}
\bibinfo{author}{S.~Bhagavantam}, \bibinfo{author}{J.~Bhimasenachar},
\newblock \bibinfo{title}{Elastic constants of {Diamond}},
\newblock \bibinfo{journal}{Nature} \bibinfo{volume}{154}
  (\bibinfo{year}{1944}) \bibinfo{pages}{546--546}.
\bibitem[{Hassle and Sudook(2001)}]{elements}
\bibinfo{author}{L.~Hassle}, \bibinfo{author}{K.~Sudook},
\newblock \bibinfo{title}{Monocrystal elastic constansts and derived properties
  of the cubic and the hexagonal elements},
\newblock in: \bibinfo{editor}{L.~Moises}, \bibinfo{editor}{B.~Henry, E.},
  \bibinfo{editor}{S.~Richard, R.} (Eds.), \bibinfo{booktitle}{Handbook of
  Elastic Properties of Solids, Liquids, and Gases, Four-Volume Set},
  volume~\bibinfo{volume}{2}, \bibinfo{publisher}{Elsvier},
  \bibinfo{year}{2001}, pp. \bibinfo{pages}{97--106}.
\bibitem[{Voigt(1910)}]{voigt1910lehrbuch}
\bibinfo{author}{W.~Voigt}, \bibinfo{title}{Lehrbuch der kristallphysik:(mit
  ausschluss der kristalloptik)}, volume~\bibinfo{volume}{34},
  \bibinfo{publisher}{BG Teubner}, \bibinfo{year}{1910}.
\bibitem[{Reuss(1929)}]{ruess1929}
\bibinfo{author}{A.~Reuss},
\newblock \bibinfo{title}{{Berechnung der Fließgrenze von Mischkristallen auf
  Grund der Plastizitätsbedingung für Einkristalle.}},
\newblock \bibinfo{journal}{ZAMM - Journal of Applied Mathematics and Mechanics
  / Zeitschrift für Angewandte Mathematik und Mechanik} \bibinfo{volume}{9}
  (\bibinfo{year}{1929}) \bibinfo{pages}{49--58}.
\bibitem[{Hill(1952)}]{Hill_1952}
\bibinfo{author}{R.~Hill},
\newblock \bibinfo{title}{The elastic behaviour of a crystalline aggregate},
\newblock \bibinfo{journal}{Proceedings of the Physical Society. Section A}
  \bibinfo{volume}{65} (\bibinfo{year}{1952}) \bibinfo{pages}{349--354}.
\bibitem[{Hill(1963)}]{HILL1963}
\bibinfo{author}{R.~Hill},
\newblock \bibinfo{title}{Elastic properties of reinforced solids: Some
  theoretical principles},
\newblock \bibinfo{journal}{Journal of the Mechanics and Physics of Solids}
  \bibinfo{volume}{11} (\bibinfo{year}{1963}) \bibinfo{pages}{357 -- 372}.
\bibitem[{İyigör and Uğur(2014)}]{elastic_Iyigor_BGValue}
\bibinfo{author}{A.~İyigör}, \bibinfo{author}{S.~Uğur},
\newblock \bibinfo{title}{Elastic and phonon properties of quaternary {Heusler}
  alloys {CoFeCrZ (Z = Al, Si, Ga and Ge)} from density functional theory},
\newblock \bibinfo{journal}{Philosophical Magazine Letters}
  \bibinfo{volume}{94} (\bibinfo{year}{2014}) \bibinfo{pages}{708--715}.
\bibitem[{Seh and Gupta(2019)}]{elastic_Seh}
\bibinfo{author}{A.~Q. Seh}, \bibinfo{author}{D.~C. Gupta},
\newblock \bibinfo{title}{Exploration of highly correlated co-based quaternary
  {Heusler} alloys for spintronics and thermoelectric applications},
\newblock \bibinfo{journal}{International Journal of Energy Research}
  \bibinfo{volume}{43} (\bibinfo{year}{2019}) \bibinfo{pages}{8864--8877}.
\bibitem[{Wu et~al.(2019)Wu, Fecher, Shahab~Naghavi, and
  Felser}]{elastic_claudia}
\bibinfo{author}{S.-C. Wu}, \bibinfo{author}{G.~H. Fecher},
  \bibinfo{author}{S.~Shahab~Naghavi}, \bibinfo{author}{C.~Felser},
\newblock \bibinfo{title}{Elastic properties and stability of {Heusler}
  compounds: Cubic {Co$_2$YZ} compounds with {L21} structure},
\newblock \bibinfo{journal}{Journal of Applied Physics} \bibinfo{volume}{125}
  (\bibinfo{year}{2019}) \bibinfo{pages}{082523}.
\bibitem[{Pettifor(1992)}]{pettifor}
\bibinfo{author}{D.~G. Pettifor},
\newblock \bibinfo{title}{Theoretical predictions of structure and related
  properties of intermetallics},
\newblock \bibinfo{journal}{Materials Science and Technology}
  \bibinfo{volume}{8} (\bibinfo{year}{1992}) \bibinfo{pages}{345--349}.
\bibitem[{Dhakal et~al.(2020)Dhakal, Nepal, Ray, Paudel, and Kaphle}]{our}
\bibinfo{author}{R.~Dhakal}, \bibinfo{author}{S.~Nepal},
  \bibinfo{author}{R.~Ray}, \bibinfo{author}{R.~Paudel},
  \bibinfo{author}{G.~Kaphle},
\newblock \bibinfo{title}{Effect of doping on sgs and weak half-metallic
  properties of {Heusler} alloys},
\newblock \bibinfo{journal}{Journal of Magnetism and Magnetic Materials}
  \bibinfo{volume}{503} (\bibinfo{year}{2020}) \bibinfo{pages}{166588}.
\bibitem[{Nepal et~al.(2020)Nepal, Dhakal, and Galanakis}]{our2}
\bibinfo{author}{S.~Nepal}, \bibinfo{author}{R.~Dhakal},
  \bibinfo{author}{I.~Galanakis},
\newblock \bibinfo{title}{Ab initio study of the half-metallic full-heulser
  compounds {Co$_2$ZAl[Z = Sc, Ti,V, Cr, Mn, Fe]}; the role of electronic
  correlations},
\newblock \bibinfo{journal}{Materials Today Communications}
  \bibinfo{volume}{25} (\bibinfo{year}{2020}) \bibinfo{pages}{101498}.
\bibitem[{Jiles(1991)}]{jiles2015introduction}
\bibinfo{author}{D.~Jiles}, \bibinfo{title}{Introduction to magnetism and
  magnetic materials}, \bibinfo{publisher}{CRC press}, \bibinfo{year}{1991}.

\end{thebibliography}

\end{document}